%% file: colm2026_conference.tex
\documentclass{article} 
\usepackage[preprint]{colm2026_conference}

\usepackage{microtype}
\usepackage{hyperref}
\usepackage{url}
\usepackage{booktabs}
\usepackage{xcolor}
\usepackage{amsmath}
\usepackage{multirow}
\usepackage{amssymb}   
\usepackage{pifont}
\newcommand{\cmark}{\ding{51}}                     \newcommand{\xmark}{\ding{55}} 
\usepackage{graphicx}
\usepackage{array}
\usepackage{enumitem}
\usepackage{bigdelim}
\usepackage{makecell}
\usepackage[table]{xcolor}
\usepackage{algorithm}
\usepackage{algpseudocode}

\newcolumntype{L}[1]{>{\raggedright\arraybackslash}p{#1}}

\usepackage{lineno}

\definecolor{darkblue}{rgb}{0, 0, 0.5}
\definecolor{committed}{RGB}{220, 240, 220}
\definecolor{rolledback}{RGB}{255, 230, 230}
\definecolor{baseline}{RGB}{230, 230, 255}
\definecolor{neutral}{RGB}{245, 245, 230}
\hypersetup{colorlinks=true, citecolor=darkblue, linkcolor=darkblue, urlcolor=darkblue}

\title{The Art of Building Verifiers for Computer Use Agents}

\author{Corby Rosset$^{1}$\thanks{Corresponding author.}, \quad Pratyusha Sharma$^{1}$, \quad Andrew Zhao$^{1}$, \\
\textbf{Miguel Gonzalez-Fernandez$^{2}$, \quad Ahmed Awadallah$^{1}$} \\[0.5em]
$^{1}$Microsoft Research \quad $^{2}$Browserbase \\
\texttt{\{corbyrosset,pratysharma,andrewzhao,ahmed.awadallah\}@microsoft.com}
}

\begin{document}

\ifcolmsubmission
\linenumbers
\fi

\maketitle

\begin{abstract}
Verifying the success of computer use agent (CUA) trajectories is a critical challenge: without reliable verification, neither evaluation nor training signal can be trusted. In this paper, we present lessons learned from building a best-in-class verifier for web tasks we call the Universal Verifier. 
We design the Universal Verifier around four key principles: 1) constructing rubrics with meaningful, non-overlapping criteria to reduce noise; 2) separating process and outcome rewards that yield complementary signals, capturing cases where an agent follows the right steps but gets blocked or succeeds through an unexpected path; 3) distinguishing between controllable and uncontrollable failures scored via a cascading-error-free strategy for finer-grained failure understanding; and 4) a divide-and-conquer context management scheme that attends to all screenshots in a trajectory, improving reliability on longer task horizons.
We validate these findings on CUAVerifierBench, a new set of CUA trajectories with both process and outcome human labels, showing that our Universal Verifier agrees with humans as often as humans agree with each other. We report a reduction in false positive rates to near zero compared to baselines like WebVoyager ($\geq$ 45\%) and WebJudge ($\geq$ 22\%). We emphasize that these gains stem from the cumulative effect of the design choices above. We also find that an auto-research agent achieves 70\% of expert quality in 5\% of the time, but fails to discover all strategies required to replicate the Universal Verifier. We open-source our Universal Verifier system along with CUAVerifierBench\footnote{Code and Data will be available at \url{https://github.com/microsoft/fara}}.

\end{abstract}
\input{Introduction}
\input{RelatedWork}
\input{Method}
\input{Experiments}

\input{Results}
\input{Conclusion}
\clearpage
\section{Ethics Statement}
We disclose that we contracted human annotators via an external firm Browserbase, and they represented to us that those annotators were paid more than minimum wage applicable under local law. We also represent that some annotators gave us express written permission to quote  qualitative feedback they gave us about their experience judging the tasks. We do no disclose any personally identifiable information about the judges. We did not give the judges any psychologically harmful, offensive, or adult-natured tasks. 

Additionally, we disclose that parts of this work were produced by generative AI, including but not limited to auto-research studies, results, analysis, and code. We performed our best effort to verify the results were not hallucinated.




\bibliography{colm2026_conference}
\bibliographystyle{colm2026_conference}

\appendix
\include{Appendix}

\end{document}

%% file: Introduction.tex
\begin{figure}[h!]
\centering
\tiny
\includegraphics[width=0.97\textwidth]{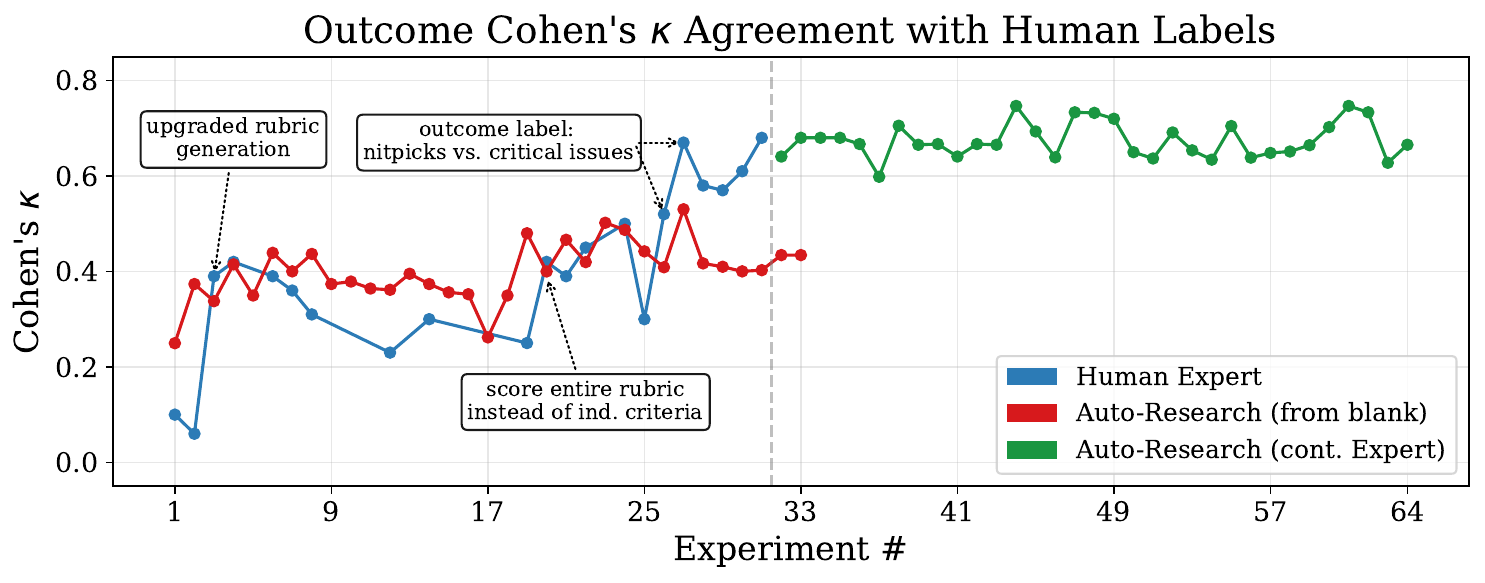}
\caption{\small We compare whether an auto-research system can design a CUA trajectory verifier as well as the expert human-designed Universal Verifier, as measured in agreement with human labels. The human expert iterated over 32 experiments across three weeks; the auto-research agent completed the same in roughly one day. Qualitatively, auto-research edits tended to be conservative and incremental, missing the design intuition behind the human's highest-impact structural decisions (tagged)}
\label{fig:kappa_progression}
\vspace{-10pt}
\end{figure}

\section{Introduction}
\vspace{-3pt}

The ability of AI agents to operate computers autonomously---browsing the web, filling forms, navigating interfaces---has advanced rapidly~\cite{zhou2024webarena, he2024webvoyager, zheng2024seeact, koh2024visualwebarena, xie2024osworld, openai2025cua, agashe2025agents2, awadallah2025fara, gupta2026molmoweb}. Yet progress in training and evaluating these systems is bottlenecked by a deceptively difficult question: \emph{did the agent actually succeed?} Unlike text generation tasks where outputs can be compared directly, computer use trajectories are long, visually rich, and ambiguous, making human annotation both challenging and expensive. The notion of success itself is nuanced: a task may be partially completed; success may be achieved through unexpected paths; and failures may be subtle, appearing only transiently in a screenshot buried deep in a multi-step interaction. Building a verifier that reliably answers this question is far from straightforward---and the consequences of getting it wrong compound, corrupting both benchmarks and training data.

In this paper, we document the lessons learned from building a verifier for computer use agents, structured as a set of actionable design principles. Our approach rests on four core ideas. First, a good verifier requires well-designed rubrics with specific, non-overlapping criteria that enable consistent scoring across diverse tasks. Second, it must report both process and outcome rewards---these provide complementary signals that differ primarily in whether the environment prevented success despite correct agent behavior, or allowed success via an unexpected but valid path. Third, it must distinguish controllable failures from uncontrollable ones and score trajectories with a cascading-error-free rubric, so that a single early obstacle does not unfairly penalize all downstream steps. Fourth, it must attend effectively to all screenshot evidence in a trajectory, not just the most recent frames; longer tasks contain critical state changes that are systematically missed when context is truncated.

To support rigorous evaluation of these principles, we release \textbf{CUAVerifierBench}, a benchmark of human-labeled CUA trajectories. To our knowledge, CUAVerifierBench is the first benchmark designed specifically to measure verifier quality for both process and outcome rewards, enabling the community to compare verifier alignment with human judgment in a standardized way. We show that our verifier---which we call the \emph{Universal Verifier}---substantially improves alignment with human labels over existing WebJudge, WebVoyager as measured by Cohen's $\kappa$, while reducing false positive rates from 30\%+ to 1-8\%.

Crucially, building a high-quality verifier is not a one-shot problem but an iterative development process, and this process is only possible when grounded in a reliable evaluation procedure. CUAVerifierBench serves exactly this role: each candidate verifier design can be scored against human judgments using Cohen's $\kappa$, providing a clear and immediate signal for what works and what does not. Figure~\ref{fig:kappa_progression} traces this iterative journey over 96 experiments. The expert-designed verifier begins with near-zero agreement and steadily improves through principled experimentation, reaching $\kappa \approx 0.7$ by experiment~32 as the four design principles are incrementally discovered and integrated. 

We also explored whether an automated research agent could replicate this process. Starting from a blank slate, the auto-research-designed verifier follows a similar upward trend but consistently underperforms, with $\kappa$ plateauing around $0.55$---roughly 70\% of expert-level quality. Qualitatively, the auto-research agent's edits tended to be conservative and incremental, struggling to encode the kind of evaluative judgment behind the large structural changes that drove the expert-designed verifier's step-function improvements. However, when initialized from the expert's best verifier configuration, the auto-research agent surpasses the expert-designed peak, suggesting that human expertise and automated optimization play complementary roles: the former is essential for discovering core design principles, while the latter excels at the fine-grained tuning that extracts remaining performance.

In summary, our contributions are as follows: (1)~We identify and validate four design principles for building reliable CUA verifiers, showing that their cumulative effect yields a verifier that agrees with humans as often as humans agree with each other. (2)~We release \textbf{CUAVerifierBench}, the first benchmark specifically designed to evaluate verifier quality for computer use agents, providing the community with a standardized way to measure verifier alignment with human judgment.

%% file: RelatedWork.tex
\section{Background and Related Work}
Several systems have been proposed for automatically evaluating CUA, differing primarily in what inputs they consume and whether they rely on prompted LLMs or trained models. \textbf{WebVoyager}~\citep{he2024webvoyagerbuildingendtoendweb} uses a GPT-4V-based evaluator that receives all trajectory screenshots (but not all action history) alongside the agent's stated final answer to produce a binary outcome judgment. Validated against human annotations on 300 tasks, the GPT-4V variant achieves 85.3\% agreement ($\kappa{=}0.70$), matching human inter-annotator agreement. \textbf{WebJudge}~\citep{xue2025illusionprogressassessingcurrent} addresses two known failure modes of this approach: reliance on the agent's potentially hallucinated final answer, and token overload from passing all screenshots unfiltered. It employs a three-step pipeline that first extracts key points from the task description, scores each screenshot for relevance, and judges success using only the top-$k$ selected screenshots and the full action history. Under the same evaluation setting, WebJudge (o4-mini) achieves 85.7\% human-agreement compared to 78.7\% for WebVoyager.

Shifting from outcome prediction to failure diagnosis, \textbf{AgentRx}~\citep{barke2026agentrxdiagnosingaiagent} identifies the \emph{critical failure step} and assigns it a root cause from a nine-category taxonomy. 

\textbf{AgentRewardBench}~\citep{lu2025agentrewardbenchevaluatingautomaticevaluations} provides 1,302 expert-annotated trajectories across five benchmarks (WebArena~\citep{zhou2023webarena}, VisualWebArena~\citep{koh2024visualwebarena}, AssistantBench~\citep{yoran2024assistantbench}, WorkArena~\citep{drouin2024workarena}, WorkArena++~\citep{boisvert2024workarenapp}) and four agent LLMs (GPT-4o~\citep{openai2024gpt4osystemcard}, Claude 3.7 Sonnet~\citep{anthropic2025claude37sonnet}, Llama-3.3-70B~\citep{grattafiori2024llama3herdmodels}, and Qwen2.5-VL~\citep{bai2025qwen25vl}). They introduce a Simplified Judge that, in a single LLM completion, predicts three binary labels---task success, side effects, and repetition cycles.
Their key finding is that no LLM-based judge exceeds 70\% precision: including NNetNav~\citep{murty2025nnetnavunsupervisedlearningbrowser} and AER~\citep{pan2024autonomousevaluationrefinementdigital}. 
Human inter-annotator agreement was 89.3\%.

Several works debated whether process or outcome rewards are more effective for scenarios such as solving math problems~\citep{lightman2023letsverifystepstep, uesato2022solvingmathwordproblems}; ~\cite{wang2024mathshepherd} trains their own process reward model. ~\cite{zhang2025lessons} distill lessons for building process verifiers for math. Others extend to agentic RAG domains~\citep{zhang2025processoutcome}. We refer the reader to additional surveys~\citep{zheng2025survey, stuhlmuller2022supervise}.

%% file: Method.tex
\section{What is True of Good Verifiers?}
\label{sec:method}

\input{tables/judge_comparison_overview}

 We distill principles we believe are critical to the construction of a reliable verifier based on our extensive hands-on experience with CUA trajectory logs.

\subsection{Good Rubrics have Specific and Non-Overlapping Criteria}
\label{sec:building_rubrics}
The root of the pipeline is rubric generation: flawed rubrics produce errors that cascade through the pipeline and cannot be easily corrected downstream. 
Anecdotally, Figure~\ref{fig:kappa_progression} shows that the good rubric design \emph{alone} accounted for roughly half of the Cohen's $\kappa$ gains.
Through iterative development, we identified four systematic failure modes and corresponding fixes:
\begin{enumerate}[nosep,leftmargin=2em]
    \item \textbf{Phantom criteria.} LLM-generated rubrics frequently introduce requirements never stated in the task (e.g., in Appendix~\ref{sec:rubric_failure_modes}, Table~\ref{tab:old_vs_new_rubric}), inflating the denominator and over-penalizing agents that completed the actual task.
    \item \textbf{Cascading errors.} When rubric criteria are not logically independent, a single upstream error propagates into downstream criteria, multiplying the point penalty.
    \item \textbf{Separate Generation and Scoring} Generating the rubric and scoring it in a single LLM call leads the model to create criteria tailored to the agent's behavior. We separate rubric generation (from the task alone, without seeing the trajectory) from scoring.
    \item \textbf{Hallucination detection.} We score the whole rubric in two passes---with and without evidence from the relevant screenshots---to surface discrepancies.
    \item \textbf{Conditional Criteria}: Some criteria may not apply depending on reality (e.g., \emph{``buy organic blueberries, or if unavailable, buy non-organic''}). Hence, at rubric-generation time, we mark some criteria as ``conditional'' to be updated once the task is attempted. Conditions that are not met are excluded, ensuring that mutually exclusive criteria do not interfere. See Appendix~\ref{sec:rubric_failure_modes}, Table~\ref{tab:conditional_criteria} for details and examples.
\end{enumerate}
 
The remaining sub-sections discuss the scoring of rubrics. Sometimes, the rubric is modified for e.g. updating conditional criteria, or adding new criteria for unsolicited side-effects. 

\subsection{Separate Process and Outcome Rewards}
\label{sec:separating_rewards}

In computer use settings, the environment plays an out-sized role in the success of a task, especially if an agent is blocked or can't access necessary resources. Hence, a central design principle of our verification framework is the separation of \emph{how well the agent executed} in the context of the environment from \emph{whether the user's goal was achieved}. These two questions have fundamentally different answers in many real-world scenarios, and conflating them leads to reward signals that are either too lenient (crediting agents for apparent effort when the user is left empty-handed) or too harsh (penalizing agents for factors outside their control). 
We formalize this separation through two
two independent signals per trajectory: a \textbf{process reward} (a fine-grained rubric whose score reflects execution across sub-goals) and an \textbf{outcome reward} (a binary success/failure judgment on whether the goal was achieved).

\paragraph{Process Label (Rubric Score):}
This is a scored rubric of criteria, each of which is weighted by a maximum number of earnable points. It is reported as a normalized score from 0.0 to 1.0 reflecting how well the agent executed each sub-goal of the task. It is computed as:
\begin{equation}
r_\text{proc} = \frac{\sum_{i \in \mathcal{A}} \text{earned\_points}_i}{\sum_{i \in \mathcal{A}} \text{max\_points}_i}
\end{equation}
where $\mathcal{A}$ is the set of \emph{applicable} rubric criteria---those whose conditions are met (for conditional criteria) or that are unconditional. The process label evaluates the quality of the agent's execution at each step, independent of whether those steps ultimately produced a successful outcome. While it is technically a scalar score, the rubric also contains specific justifications as to why points were earned or lost based on evidence from the full action history and screenshots. 
An agent that, for example, navigated to the correct product, and but was blocked by a login wall before it could add-to-cart would full process credit, even though the user's goal was not achieved. Example rubrics can be seen in Figures~\ref{fig:example_verifier_result}, and~\ref{fig:example_verifier_result_criterion}.

\paragraph{Outcome Label (Binary Success/Failure)}
The outcome label is a binary yes/no judgment answering: \textbf{would a reasonable user consider the task done?} This is evaluated from the perspective of a user who issued the task and is examining the end state. This is intrinsically challenging, because users may have different notions of success under ambiguity (e.g. is it acceptable to omit NeurIPS's secondary venue in Mexico City when asked \emph{``where is NeurIPS 2025 being hosted?''}) and different preferences as for what constraints are strict vs flexible (e.g. is ok to book a table using opentable.com when the user asked to use resy.com?). 

In order to make progress, we assume that the outcome label should focus on \emph{primary intent} -- if the primary intent is to book a table, then the user would be flexible on which platform it is booked unless otherwise stated. We also believe most users are forgiving of nitpicks like rounding \$5.95 to \$6, etc. However, we assume users would \emph{not} be forgiving of unsolicited side-effects e.g. buying a warranty when they only wanted to buy the product itself, or hallucinations like those described in Table~\ref{tab:visual_evidence}. We summarize the process and outcome rewards for computer use scenarios in Table~\ref{tab:scenarios} -- notice they only disagree in the second row.

\subsection{Discern Controllable vs. Uncontrollable Factors}
\label{sec:controllable_uncontrollable}

Since the  main difference between a trajectory being a process success but an outcome failure involves the environment, we explicitly define which of these aspects are controllable vs. uncontrollable from the perspective of the agent. Each rubric's criteria description fields attempt to anticipate these factors and give guidance on how to award partial credit.

\textbf{Uncontrollable factors:}  Conditions beyond the agent's control; \emph{not} penalized in \emph{process}.
\begin{itemize}[nosep,leftmargin=2em]
    \item \textbf{Platform/infrastructure issues}: CAPTCHA, login walls without credentials, etc.
    \item \textbf{Entity non-existence}: product discontinued, business closed, service not available.
    \item \textbf{Availability constraints}: out of stock, no reservations on requested date, sold out.
    \item \textbf{Search result limitations}: no results matching all specified criteria.
\end{itemize}

\textbf{Controllable factors:} Avoidable mistakes the agent \emph{should} be penalized for in \emph{process}.
\begin{itemize}[nosep,leftmargin=2em]
    \item \textbf{Intent Mis-match}: Choosing an entirely wrong product, location, person, service, etc.
    \item \textbf{Reasoning Errors}: Incorrect reasoning about the task e.g. Figure~\ref{fig:controllable_error_example}.
    \item \textbf{Hallucinations}: claiming success without evidence, fabricating information.
    \item \textbf{Insufficient effort}: giving up after a single failed attempt.
    \item \textbf{Execution errors}: not using available filters, skipping required steps.
\end{itemize}

\subsection{Effective Context Management of Screenshot Evidence}

Our main contribution is a verifier designed to combat hallucinations\footnote{We define the anatomy of hallucinations Section~\ref{sec:visual_evidence}, Table~\ref{tab:visual_evidence}, and give an example in Figure~\ref{fig:hallucination}} through better management of visual screenshot evidence. Both WebVoyager~\citep{he2024webvoyagerbuildingendtoendweb} and WebJudge~\citep{xue2025illusionprogressassessingcurrent} assess a large amount of screenshots in one LLM context window -- WebVoyager includes all screenshots, whereas WebJudge ranks the top $~\approx 30-50$. Other verifiers only analyze the last ones~\citep{pan2024autonomousevaluationrefinementdigital}.  Too many screenshots over-exerts the LLM by forcing it to solve a needle-in-a-haystack problem, which scales poorly with longer trajectories, whereas  restricting to the last few risks missing task-relevant evidence. 
To address these problems, our design scores each screenshot against every rubric criterion to produce a relevance matrix, grouping the top-$k$ most relevant \emph{per criterion} to send for further analysis, which is both more scalable to longer trajectories and more focused. We elaborate on our screenshot-scoring design in Appendix~\ref{sec:screenshot_rel} with an example in Figure~\ref{fig:relevance_matrix}.

\subsection{Unsolicited Side-Effects}  Extraneous actions with material side effects—such as adding unrequested items to a cart (e.g. see Figure~\ref{fig:unsolicited_side_effect}) or enrolling in unrequested services—constitute often cannot be anticipated before the task is attempted because rubrics are not designed to enumerate all the ways a task can go wrong. To catch such cases, a dedicated pass over a trajectory is needed.  While unsolicited side-effects almost always result in outcome failure, they only partially penalize the process score, weighted by how serious the side-effect is. 

\section{Universal Verifier System}
\label{sec:pipeline_overview}

We model a computer use task as a tuple $(g, \mathcal{E})$, where $g$ is a natural language goal (e.g., \textit{``book the cheapest available flight from Seattle to Boston on June 3rd''}) and $\mathcal{E}$ is a computer environment with an observable graphical interface. An agent interacts with $\mathcal{E}$ over $T$ discrete timesteps, producing a trajectory $\tau = (s_0, a_1, s_1, a_2, \ldots, a_T, s_T)$, where $s_t \in \mathcal{S}$ is a screenshot observation at time $t$ and $a_t \in \mathcal{A}$ is an action (e.g., click, type, scroll). The length $T$ varies across tasks from a handful of steps for form-filling to hundreds of steps for multi-stage workflows.

We define a \textbf{verifier} as a function $V: (g, \tau) \to \mathcal{R}$ that maps a goal and trajectory to a structured scoring response $r \in \mathcal{R}$. In the simplest case $\mathcal{R} = \{0, 1\}$ (binary success), but we argue and our design reflects---that $\mathcal{R}$ should be richer: a tuple $(r_\text{proc}, r_\text{out}, d)$ comprising a process score $r_\text{proc} \in [0,1]$, an outcome score $r_\text{out} \in \{0, 1\}$, and a diagnostic report $d$ that classifies and localizes failures within $\tau$. The process score captures the quality of the agent's execution, while the outcome score reflects whether the goal $g$ was ultimately satisfied. 

The central challenge is that $V$ must operate over the full observation sequence $\{s_0, \ldots, s_T\}$, which can be long, visually dense, and contain critical state changes at arbitrary timesteps. We define verifier quality as agreement with a human oracle $V^*: (g, \tau) \to \mathcal{R}$, measured by precision, recall, and Cohen's $\kappa$ over a labeled set of trajectories~\citep{artstein-poesio-2008-survey}. A verifier that inspects only $s_T$ or a fixed subset $\{s_{t_1}, \ldots, s_{t_k}\} \subset \tau$ is a strict approximation of $V^*$ and, as we show empirically, systematically underperforms on trajectories where $T$ is large. Reliable verification therefore requires attending to all $T+1$ observations.

\begin{algorithm}[t]
\small
\caption{Universal Verifier}
\label{alg:eval_pipeline}
\begin{algorithmic}[1]
\Require agent trajectory $\tau$, observations $\{s_0, \dots, s_T\}$, user goal $g$
\Ensure Process score $r_{\text{proc}}$, Outcome score $r_{\text{out}}$, diagnostic report $d$
\State \textbf{Generate Rubric}
    $\mathcal{C} = \{c_1, \dots, c_N\}$ of $N$ disjoint, meaningful criteria from $g$. See \ref{sec:rubric_failure_modes}
\State \textbf{Multimodal Relevance Scoring.}
    Score each screenshot against every criterion to produce relevance matrix
    $\mathbf{R} \in \mathbb{R}^{(T+1) \times N}$. See Appendix~\ref{sec:screenshot_rel} for more details.
\State \textbf{Top-$k$ Grouping.}
    For each $c_j$, select the $k$ most relevant
    $\mathcal{S}_j \subseteq \{s_0,\dots,s_T\},\; |\mathcal{S}_j| \leq k$.
\State \textbf{Evidence Analysis.}
    For each pair $(c_j,\, s_i)$ with $s_i \in \mathcal{S}_j$, extract visual evidence $e_{ij}$.
\State \textbf{Conditional Disambiguation.}
    Resolve conflicts among conditional criteria using $\{e_{ij}\}$.
\State \textbf{Reality Check.}
    Reconcile rubric assumptions against screenshot evidence;
    produce interpretive reality notes and action-only score $r_{\text{proc\_action\_only}}$.
\State \textbf{Multimodal Rescoring.}
    Rescore $\mathcal{C}$ holistically using screenshot evidence (which takes precedence over agent claims) following Tables \ref{tab:scenarios} and \ref{tab:visual_evidence}.
\State \textbf{Side-Effect Detection.}
    Detect and include unsolicited agent actions with material side effects not already penalized by $\mathcal{C}$, \Return procedural score $r_{\text{proc}}$. (see example Figure~\ref{fig:unsolicited_side_effect})
\State \textbf{Outcome Verification.}
    run and \Return outcome score $r_{\text{out}}$.
\State \textbf{Failure Diagnosis.}
    Identify and localize all failures points from Table~\ref{tab:error-taxonomy} and \Return
    $d$.
\end{algorithmic}
\end{algorithm}

The \textbf{Universal Verifier} (UV) we create incorporates the principles from Section~\ref{sec:method} and operates in three phases: rubric creation, \emph{multimodal scoring} incorporating screenshot evidence to ascertain $r_\text{proc}$, and produce a final outcome judgment $r_\text{out}$, and error diagnosis $d$ as shown in Algorithm~\ref{alg:eval_pipeline}.
The key design invariant is that no relevant screenshot evidence can go undetected in the pipeline, specifically to not miss any hallucinations.
To reduce variance, Steps 7--9 in Algorithm~\ref{alg:eval_pipeline} can be run as multiple parallel instances, with process score determined by median of rubric scores, and outcome by majority vote. 

Finally, we conduct an error analysis on $\tau$ to categorize failure modes and identify the step $t$ at which each failure occurred in a trajectory. We hand-crafted an error taxonomy with 7 categories and 24 subcodes as shown in Table~\ref{tab:error-taxonomy}, covering categories such as intent missmatches, hallucinations, critical point violations, etc.

%% file: tables/judge_comparison_overview.tex
\begin{table*}[t]
\vspace{-10pt}
\centering
\small
\setlength{\tabcolsep}{4pt}
\begin{tabular}{l l l l l l}
\toprule
\textbf{Verifier} & \textbf{LLM} & \textbf{Rubric} & \textbf{Screenshots} & \textbf{Action hist.} & \textbf{Final ans.} \\
\midrule
\multirow{3}{*}{\shortstack[l]{WebJudge\\(OM2W)}} & \multirow{3}{*}{o4-mini} & \multirow{3}{*}{\textcolor{red!70!black}{\xmark}~Not used} & \textcolor{green!70!black}{\cmark}~Top-$k$ most relevant & \multirow{3}{*}{\textcolor{green!70!black}{\cmark}~Full} & \multirow{3}{*}{\textcolor{red!70!black}{\xmark}} \\
 & & & (scored 1--5, kept if & & \\
 & & & $\geq$ threshold; capped at 5) & & \\
\midrule
\multirow{3}{*}{\shortstack[l]{WebVoyager\\GPT eval}} & \multirow{3}{*}{gpt-4o} & \multirow{3}{*}{\textcolor{red!70!black}{\xmark}~Not used} & \textcolor{green!70!black}{\cmark}~All screenshots & \multirow{3}{*}{\textcolor{red!70!black}{\xmark}} & \multirow{3}{*}{\textcolor{green!70!black}{\cmark}} \\
 & & & (last $N$ if over limit; & & \\
 & & & default $N{=}30$) & & \\
\midrule
\multirow{2}{*}{\shortstack[l]{Universal\\Verifier (Ours)}} & \multirow{2}{*}{gpt-5.2} & \textcolor{green!70!black}{\cmark}~Per-task & \textcolor{green!70!black}{\cmark}~Top-$k$ most & \multirow{2}{*}{\textcolor{green!70!black}{\cmark}~Full} & \multirow{2}{*}{\textcolor{green!70!black}{\cmark}} \\
 & & success criteria & relevant \emph{per criterion} & & \\
\bottomrule
\end{tabular}
\caption{\small Comparison of different computer use trajectory verifiers' characteristics}
\label{tab:judge-comparison}
\vspace{-10pt}
\end{table*}

%% file: Experiments.tex
\section{Experiments}
\label{sec:experiments}
We treat the Universal Verifier as an annotator like any other human, and compute inter-annotator agreements throughout our studies: (1)~agreement with human trajectory labels on two independently annotated datasets, (2)~agreement between native benchmark verifiers and UV at scale, and (3)~an auto-research study exploring whether an AI agent can replace or augment human expertise in verifier design. We describe each experimental setup below.

\subsection*{CUAVerifierBench: Human-Labeled Datasets}
\label{sec:human_datasets}

Since the UV's innovation of verifying both process and outcome labels is novel in the computer use domain, no existing benchmarks provide both labels. 
 
We sampled 140 trajectories from WebTailBench using Fara-7B\citep{awadallah2025fara7b}. In-house expert annotators labeled each trajectory for both process success and outcome success following the guidelines in \S\ref{sec:method}. This dataset is used for all ablation studies (\S\ref{sec:main_comparison}--\ref{sec:fixed_rubric}) and the auto-research experiments (\S\ref{sec:auto_research}). We call this the \textbf{Internal dataset}.

Furthermore, we contracted external annotators managed by \textbf{Browserbase}\footnote{\url{https://www.browserbase.com/}} to label 106 trajectories sampled from Fara-7B \citep{awadallah2025fara7b} on Online-Mind2Web for both process and outcome success, with $2\times$ annotator overlap per trajectory. Annotators were first calibrated on 10 practice trajectories with gold annotations. They then judged each evaluation trajectory in a two-stage process:
1) \textbf{UV-blind stage:} Annotators saw only the input task, the un-scored rubric criteria, and the agent's trajectory. They independently judged outcome and process success and provided a continuous rubric score per trajectory.
2) \textbf{UV-informed stage:} Annotators were  shown the UV's outcome verdict and rubric scores, and asked whether they \emph{agreed}/\emph{disagreed} with the UV's outcome and process. 

For task-level aggregation, outcome labels are computed as the majority vote of the annotators' binary judgments, and process labels are the median of the annotators' continuous rubric scores, then binarized at a $\geq 0.8$ threshold. Ties are broken by a third.
We report agreement metrics from both stages: \emph{UV-blind} agreement measures how often human judgments independently align with the UV, while \emph{UV-informed} agreement measures how often humans endorse the UV's verdict after reviewing its reasoning. We further measure inter-annotator agreement, and how often their labels flipped once seeing the UV's output.

\paragraph{Agreement on Canonical Benchmarks' Verifiers}
\label{sec:native_agreement_setup}

The human-labeled datasets above are small by design (expert annotation is expensive). To assess verifier behavior at scale, we re-score several agent trajectories across several canonical benchmarks like with Universal Verifier and compute agreement between that benchmark's ``native'' verifier and UV. We select three benchmarks -- WebVoyager, Online-Mind2Web (OM2W), and WebTailBench -- and two agent models -- Fara-7B and GPT-5 as a Set-of-Marks Agent~\citep{yang2023setofmarkpromptingunleashesextraordinary}. 



\paragraph{Auto-Research Study}
\label{sec:auto_research_setup}
The Universal Verifier comprises approximately 3{,}000 lines of code and 2{,}000 lines of prompts---including rubric generation templates, scoring instructions, outcome verification logic, and error classification rules---all designed iteratively by a human expert (the first author). To investigate whether an AI agent can replicate or augment this human expertise, we designed an auto-research system using Claude Code v2.1.87 with Claude Opus 4.6 (1M context) on a Claude Max subscription. The system is given the same principles from Section~\ref{sec:method}, and reuses the same experimental infrastructure as the human expert (running the UV on the internal set, computing agreement metrics, and committing prompt changes to version control). We evaluate two settings:

\begin{itemize}[nosep,leftmargin=2em]
    \item \textbf{From-blank prompts:} All ${\sim}2{,}000$ lines of prompts are replaced with \texttt{// TODO} placeholders, leaving only the code scaffold. The agent is given high-level design principles but no access to prior prompt versions, previous commits, or other branches. A separate compliance agent audits each iteration to prevent memorization of test examples into prompts. The optimization rule is: \emph{maximize Cohen's $\kappa$ without increasing FPR}; any FPR-increasing change is automatically rolled back.
    \item \textbf{Continuing expert work:} The agent starts from the human expert's best prompts and continues with the same optimization objective.
\end{itemize}

%% file: Results.tex
\section{Results}
\label{sec:results}

\textbf{Agreement with Human Labels: UV vs. Existing Verifiers}: In Table~\ref{tab:main_verifier_comparison} we compare UV against two prominent existing trajectory judges---WebVoyager \citep{he2024webvoyagerbuildingendtoendweb} and WebJudge \citep{xue2025illusionprogressassessingcurrent}---on CUAVerifierBench. The UV substantially outperforms both baselines across nearly every metric on both datasets. On outcome labels, the UV achieves a Cohen's $\kappa$ of 0.64 (internal) and 0.58 (Browserbase), compared to 0.44/0.26 for WebJudge and 0.31/0.13 for WebVoyager. Strikingly, the UV achieves an FPR near zero (0.01 internal, 0.08 Browserbase) on outcome labels, meaning it almost never credits a trajectory with success when a human annotator would mark it as a failure. A version of this table with standard deviation error bars computed from three independent runs is included in Table~\ref{tab:extended_main_verifier_comparison}.
\label{sec:main_comparison}

To test whether the UV's advantage stems from simply from using a stronger backbone model, we report four additional columns in Table~\ref{tab:main_verifier_comparison}, where we upgrade WebVoyager's GPT-4o and WebJudge's o4-mini to GPT-5.2. While this does reduce FPR substantially (e.g., WebVoyager outcome FPR drops from 0.45 to 0.10 on Internal), it also dramatically increases FNR (0.24$\to$0.44), and overall $\kappa$ improves only modestly. \textbf{We conclude UV's advantage stems from its screenshot scoring design, not merely from using a stronger model}

\input{tables/main-table}

\textbf{Browserbase Annotations:}
Using the two-stage annotation protocol described in \S\ref{sec:human_datasets}, we measure how agreement changes when annotators are shown the UV's reasoning. The UV-informed stage substantially improves agreement: outcome Cohen's $\kappa$ rises from 0.39 to 0.63, and outcome FNR drops from 0.62 to 0.35, while FPR remains near zero (0.04). On process labels, FNR drops sharply from 0.32 to 0.09. Only 16.6\% of annotator outcome judgements flipped after seeing the UV's reasoning, nearly all moving from success to failure after the UV identified a failure they initially missed. 

We also plot a scatter plot of the rubric score the human annotators assigned to the trajectories vs what the UV assigned in Figure~\ref{fig:rubric_scatter}. See in Appendix~\ref{sec:browserbase_results} including Table~\ref{tab:browserbase_human_agreement} for full results.

\textbf{Inter-annotator agreement:} the Browserbase split contains at least two annotations per trajectory. The UV's outcome $\kappa$ with human labels (0.58, Table~\ref{tab:main_verifier_comparison}) and process $\kappa$ (0.43) fall within the corresponding inter-annotator ranges (0.53--0.57 and 0.36--0.45, respectively; Table~\ref{tab:inter_annotator}), \textbf{indicating that the UV agrees with humans about as well as humans agree with each other on both dimensions} (We report more details in Section~\ref{sec:browserbase_results}).

\textbf{Ablations: Varying Rubric Generator and Scorer}: We conduct two additional ablations of the Universal Verifier, reported in full in Appendix~\ref{sec:ablation_appendix}. In Table~\ref{tab:alignment_various_llms} we vary the backbone LLMs of the UV end-to-end (each model generates and scores its own rubric), finding that GPT-5.2 achieves the lowest FPR while GPT-5 offers the best balanced agreement. In Table~\ref{tab:alignment_fixed_rubric} we again vary the backbone LLM, but isolate the scoring component by fixing the rubric (generated by GPT-5.2), showing that GPT-5.2 is the most conservative scorer while GPT-5.1 achieves the highest overall $\kappa$.
\label{sec:varying_rubric}
\label{sec:fixed_rubric}

\textbf{Agreement Between UV and Native Benchmark Verifiers
}We measure agreement between the UV and the \emph{native verifiers} shipped with each of three benchmarks: WebVoyager, Online-Mind2Web (OM2W), and WebTailBench. Table~\ref{tab:native_agreement} shows that the native verifiers disagree substantially with the UV labels: false positive rates w.r.t UV outcome labels are consistently above 20\%, with WebVoyager (GPT-4o) having the highest FPR and lowest Cohen's $\kappa$. Histograms of error taxonomies for these are shown in Figures~\ref{fig:webvoyager-error-histogram}, ~\ref{fig:om2w-error-histogram}, and ~\ref{fig:webtailbench-error-histogram}.
\label{sec:native_agreement}

\input{tables/native_verifier_agreement}

\paragraph{Auto-Research: Can AI Replace Human Experts in Verifier Design?}
\label{sec:auto_research}

A natural question is whether an AI auto-research agent can replicate---or even improve upon---the process of designing verifiers\citep{lu2026towards, karpathy2026autoresearch, tie2025surveyaiscientists}. 
Figure~\ref{fig:kappa_progression} shows outcome Cohen's $\kappa$ progression across experiments for the human expert and both auto-research settings (process $\kappa$ is in Figure~\ref{fig:kappa_progression_process}), and Figures~\ref{fig:fpr_fnr_progression_outcome}--\ref{fig:fpr_fnr_progression_process} show the corresponding FPR and FNR trajectories.
The blank-prompt auto-research agent reached about ~70\% of the quality of the human expert in only 5\% of the time, and when given the best prompts and code the human had, it could still find improvements subject to the constraint of not increasing false positive rate. 
Table~\ref{tab:autoresearch_summary} in Appendix ~\ref{sec:appendix-auto-research} summarizes each \texttt{continue-expert} iteration's purpose and whether it was committed or rolled back.

Regarding AgentRewardBench~\citep{lu2025agentrewardbenchevaluatingautomaticevaluations}, in Appendix~\ref{sec:agentrewardbench_appendix} we report that out of a sample of 30 trajectories that terminated within step budget and were labeled as successful by their human annotators, we consider 8 to be false positive according to our outcome guidelines (FPR$\approx0.27$). 

%% file: tables/main-table.tex
\begin{table*}[t]
    \centering
    \small
    \setlength{\tabcolsep}{4pt}
    \begin{tabular}{@{}l cc cc c cc cc c@{}}
    \toprule
    & \multicolumn{5}{c}{\textbf{Internal Dataset} ($n{=}140$)} & \multicolumn{5}{c}{\textbf{Browserbase OM2W} ($n{=}106$)} \\
    \cmidrule(lr){2-6} \cmidrule(lr){7-11}
    & \multicolumn{2}{c}{\textbf{WebVoy.}} & \multicolumn{2}{c}{\textbf{WebJudge}} & \textbf{UV} & \multicolumn{2}{c}{\textbf{WebVoy.}} & \multicolumn{2}{c}{\textbf{WebJudge}} & \textbf{UV} \\
    \cmidrule(lr){2-3} \cmidrule(lr){4-5} \cmidrule(lr){6-6} \cmidrule(lr){7-8} \cmidrule(lr){9-10} \cmidrule(lr){11-11}
    & {\scriptsize GPT-4o} & {\scriptsize GPT-5.2} & {\scriptsize o4-mini} & {\scriptsize GPT-5.2} & {\scriptsize GPT-5.2} & {\scriptsize GPT-4o} & {\scriptsize GPT-5.2} & {\scriptsize o4-mini} & {\scriptsize GPT-5.2} & {\scriptsize GPT-5.2} \\
    \midrule
    \multicolumn{6}{@{}l|}{\textit{Agreement with outcome human labels}} & \multicolumn{5}{l}{} \\
    \quad Accuracy ($\uparrow$)            & $0.67$ & $0.70$ & $0.72$ & $0.64$ & \multicolumn{1}{c|}{${0.81}$} & $0.48$ & $0.74$ & $0.64$ & $0.74$ & ${0.88}$ \\
    \quad F1 ($\uparrow$)                  & $0.73$ & $0.69$ & $0.74$ & $0.58$ & \multicolumn{1}{c|}{${0.81}$} & $0.35$ & $0.50$ & $0.44$ & $0.46$ & ${0.65}$ \\
    \quad Cohen's $\kappa$ ($\uparrow$)    & $0.31$ & $0.43$ & $0.44$ & $0.33$ & \multicolumn{1}{c|}{\colorbox{yellow}{$0.64$}} & $0.13$ & $0.36$ & $0.26$ & $0.31$ & \colorbox{yellow}{$0.58$} \\
    \quad FNR ($\downarrow$)               & \colorbox{yellow}{$0.24$} & $0.44$ & $0.33$ & $0.57$ & \multicolumn{1}{c|}{$0.32$} & \colorbox{yellow}{$0.12$} & $0.18$ & \colorbox{yellow}{$0.12$} & $0.29$ & $0.31$ \\
    \quad FPR ($\downarrow$)               & \textcolor{red}{$0.45$} & $0.10$ & $0.22$ & $0.07$ & \multicolumn{1}{c|}{\colorbox{yellow}{$0.01$}} & \textcolor{red}{$0.60$} & $0.28$ & \textcolor{red}{$0.40$} & $0.26$ & \colorbox{yellow}{$0.08$} \\
    \midrule
    \multicolumn{6}{@{}l|}{\textit{Agreement with process human labels}} & \multicolumn{5}{l}{} \\
    \quad Accuracy ($\uparrow$)            & $0.62$ & $0.64$ & $0.66$ & $0.61$ & \multicolumn{1}{c|}{${0.81}$} & $0.55$ & $0.75$ & $0.68$ & $0.73$ & ${0.78}$ \\
    \quad F1 ($\uparrow$)                  & $0.70$ & $0.65$ & $0.70$ & $0.57$ & \multicolumn{1}{c|}{${0.86}$} & $0.47$ & $0.56$ & $0.53$ & $0.49$ & ${0.57}$ \\
    \quad Cohen's $\kappa$ ($\uparrow$)    & $0.17$ & $0.34$ & $0.32$ & $0.30$ & \multicolumn{1}{c|}{\colorbox{yellow}{$0.59$}} & $0.22$ & $0.40$ & $0.34$ & $0.32$ & \colorbox{yellow}{$0.43$} \\
    \quad FNR ($\downarrow$)               & $0.31$ & $0.49$ & $0.40$ & $0.59$ & \multicolumn{1}{c|}{\colorbox{yellow}{$0.24$}} & \colorbox{yellow}{$0.05$} & $0.23$ & $0.12$ & $0.36$ & $0.29$ \\
    \quad FPR ($\downarrow$)               & $0.52$ & $0.10$ & $0.25$ & \colorbox{yellow}{$0.04$} & \multicolumn{1}{c|}{\colorbox{yellow}{$0.04$}} & $0.56$ & $0.26$ & $0.38$ & $0.25$ & \colorbox{yellow}{$0.20$} \\
    \bottomrule
    \end{tabular}
    \vspace{2pt}
    \caption{\small Agreement between three verifiers and humans in CUAVerifierBench. Upgrading external verifier to GPT-5.2 results in only modest improvement, confirming the UV's advantage is architectural.}
    \label{tab:main_verifier_comparison}
\end{table*}

%% file: tables/native_verifier_agreement.tex
\begin{table*}[t]
\centering
\small
\begin{tabular}{@{}l cc cc cc@{}}
\toprule
& \multicolumn{2}{c}{\textbf{WebVoyager}} & \multicolumn{2}{c}{\textbf{OM2W}} & \multicolumn{2}{c}{\textbf{WebTailBench}} \\
\cmidrule(lr){2-3} \cmidrule(lr){4-5} \cmidrule(lr){6-7}
& Fara-7B & GPT-5 & Fara-7B & GPT-5 & Fara-7B & GPT-5 \\
\midrule
$N$ (tasks scored)          & 594   & 593   & 298   & 276   & 599   & 597   \\
Unterminated (\%)           & 4.2   & 3.4   & 5.0   & 7.2   & 17.0  & 7.7   \\
\midrule
\multicolumn{7}{@{}l}{\textit{Success rate (\%)}} \\
\quad Native verifier       & 74.6  & 90.6  & 32.2  & 62.0  & 39.6  & 62.5  \\
\quad UV Process & 49.0 & 79.4 & 25.8 & 64.9 & 39.6 & 63.5 \\
\quad UV Outcome            & 37.9  & 71.0  & 15.8  & 48.6  & 23.2  & 39.9  \\
\midrule
\multicolumn{7}{@{}l}{\textit{Native vs.\ UV Process\textsuperscript{$\dagger$}}} \\
\quad FNR ($\downarrow$)    & 0.06  & 0.04  & 0.26  & 0.27  & 0.30  & 0.23  \\
\quad FPR ($\downarrow$)    & \textcolor{red}{0.56} & \textcolor{red}{0.68} & 0.18 & 0.42 & 0.20 & 0.37 \\
\quad Accuracy ($\uparrow$) & 0.69  & 0.83  & 0.80  & 0.67  & 0.76  & 0.72  \\
\quad F1 ($\uparrow$)       & 0.75  & 0.90  & 0.66  & 0.74  & 0.70  & 0.78  \\
\quad Cohen's $\kappa$ ($\uparrow$) & 0.38 & 0.36 & \textbf{0.52} & 0.30 & \textbf{0.50} & \textbf{0.40} \\
\midrule
\multicolumn{7}{@{}l}{\textit{Native vs.\ UV Outcome}} \\
\quad FNR ($\downarrow$)    & 0.01  & 0.02  & 0.17  & 0.24  & 0.14  & 0.17  \\
\quad FPR ($\downarrow$)    & \textcolor{red}{0.60} & \textcolor{red}{0.72} & 0.23 & 0.49 & 0.25 & 0.49 \\
\quad Accuracy ($\uparrow$) & 0.63  & 0.78  & 0.78  & 0.63  & 0.77  & 0.64  \\
\quad F1 ($\uparrow$)       & 0.68  & 0.86  & 0.55  & 0.67  & 0.64  & 0.65  \\
\quad Cohen's $\kappa$ ($\uparrow$) & 0.33 & 0.33 & 0.42 & 0.27 & \textbf{0.49} & 0.31 \\
\bottomrule
\end{tabular}
\vspace{2pt}

\caption{\small Agreement between native benchmark verifiers and the Universal Verifier (UV) across three benchmarks and two agent models. The UV is treated as the reference label. }
\label{tab:native_agreement}
\vspace{-10pt}
\end{table*}

%% file: Conclusion.tex
\section{Conclusion}

We presented the Universal Verifier and CUAVerifierBench, demonstrating that our four design principles cumulatively produce a verifier that 1) agrees with humans as often as humans agree with each other and 2) better than any other verifier we measured, while 3) reducing false positive rates to near zero compared to baselines like WebVoyager ($\geq$45\%) and WebJudge ($\geq$22\%). These gains are architectural rather than model-driven: upgrading baseline backbones to the same LLM used by the UV yields only modest improvements. Our auto-research experiment reveals that while an AI agent can reach 70\% of expert-level verifier quality in 5\% of the time, it struggles to independently discover the structural design decisions that drive the largest gains, suggesting that building reliable verifiers remains as much an art of encoding evaluative judgment as it is an engineering problem.

%% file: Appendix.tex
\section{Universal Verifier Details}
\subsection{Top-Level Rubric and Outcome Example}
\begin{figure}[h!]
\centering
\tiny
\includegraphics[width=\textwidth]{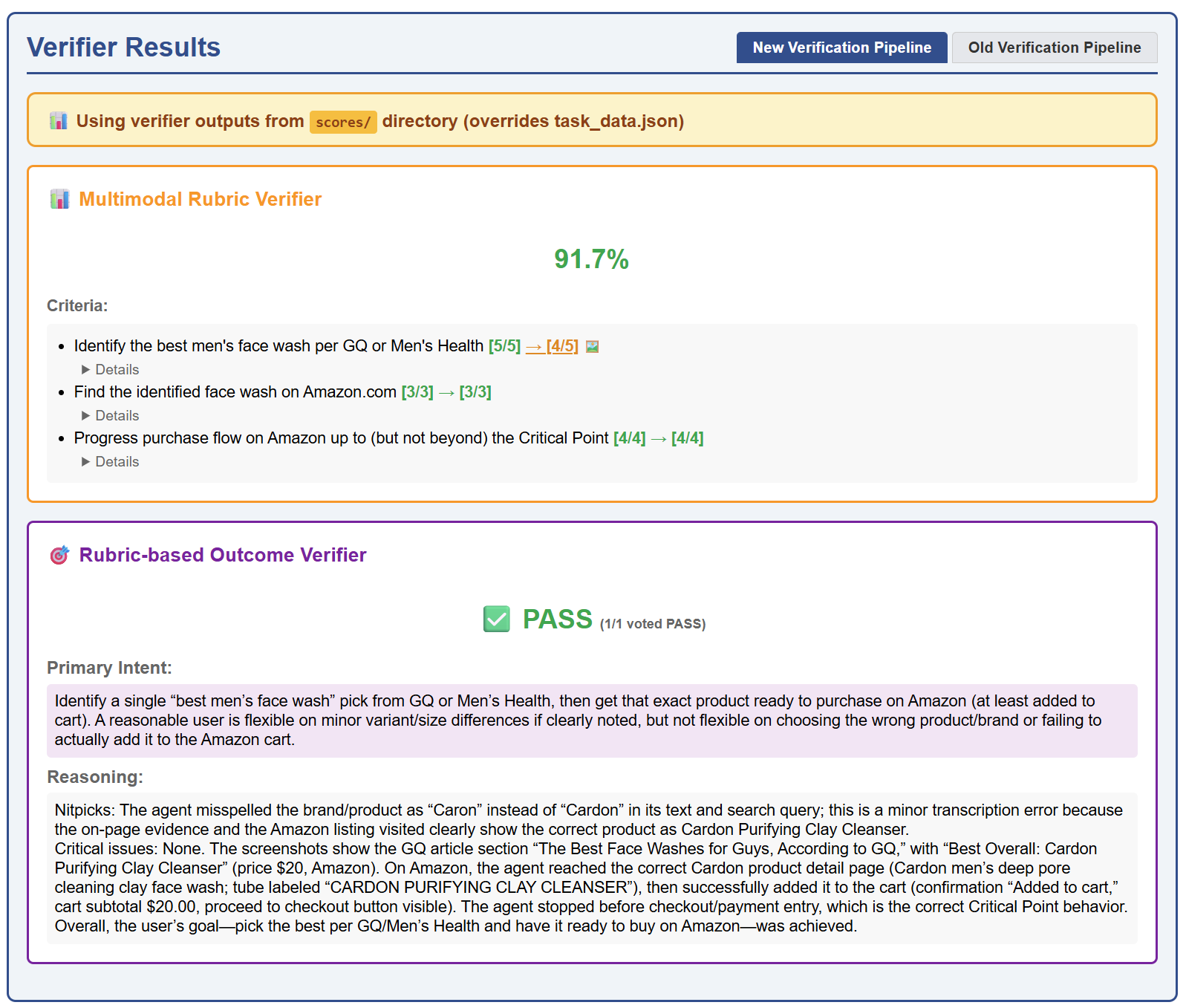}
\caption{An snapshot of our internal visualization tool for viewing verification results for a trajectory addressing the task ``find the best men's face wash according to GQ or Men's Health, then buy it on Amazon''}
\label{fig:example_verifier_result}
\end{figure}

The output of our Universal Verifier is a rubric which shows scores for individual criteria based on action-history-only scoring, which are then updated with multimodal evidence. It also shows a separate Outcome result as shown in Figure~\ref{fig:example_verifier_result}.

\begin{figure}[h!]
\centering
\tiny
\includegraphics[width=\textwidth]{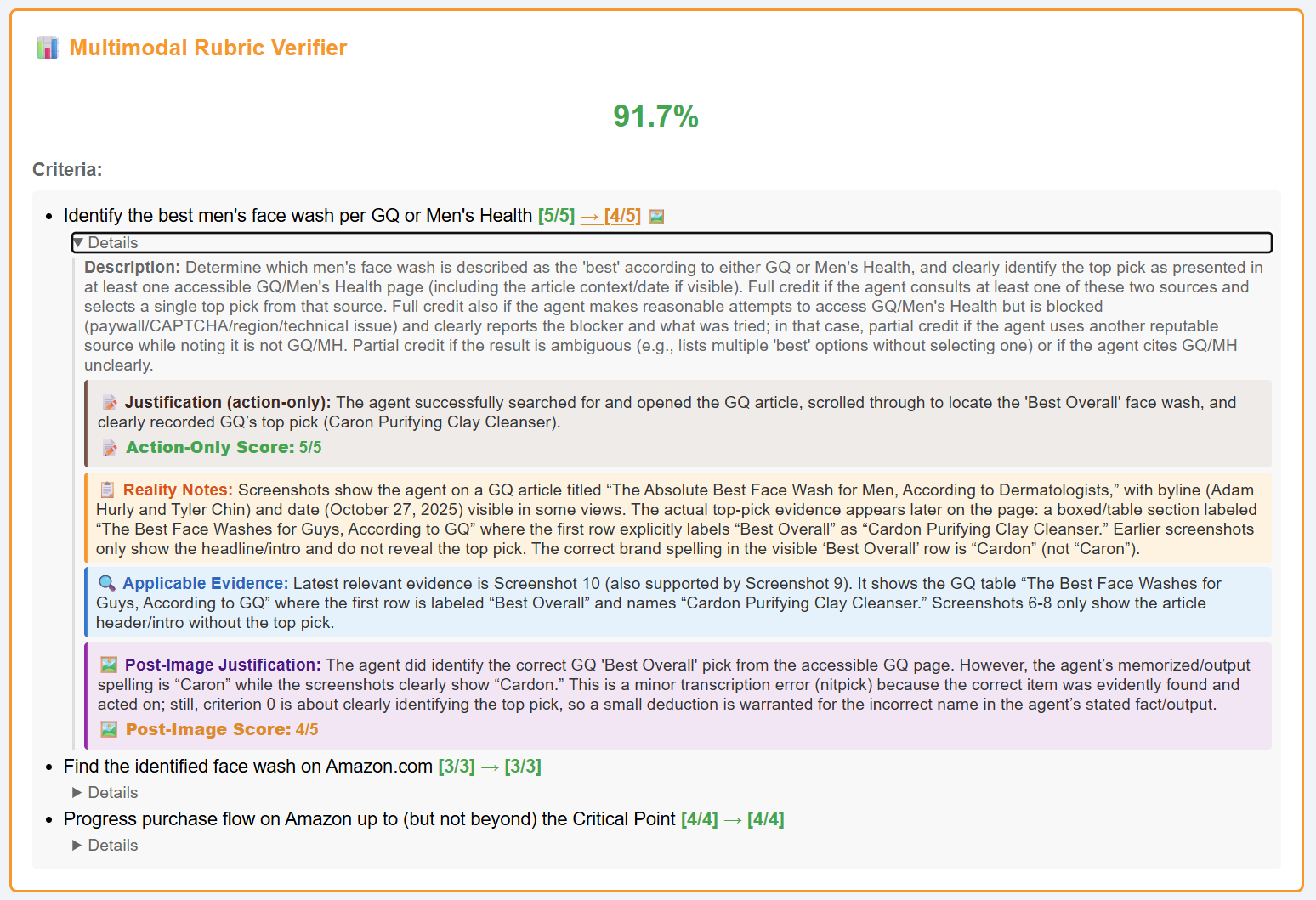}
\caption{A snapshot for the same example of how an individual criterion was scored, in this case, the model lost a point because it transcribed "Cardon" incorrectly as "Caron" in its action history based on multi-modal evidence. These kinds of meticulous analysis helps us detect hallucinations that otherwise would slip through. }
\label{fig:example_verifier_result_criterion}
\end{figure}

We record details of how each individual criterion are scored, as shown in Figure~\ref{fig:example_verifier_result_criterion}

\subsection{Rubric Failure Modes and Fixes}
\label{sec:rubric_failure_modes}

Rubric generation is the root of the verification pipeline, and flawed rubrics produce errors that cascade through scoring and outcome determination. Through iterative development (\S\ref{sec:method}), we identified several systematic failure modes in LLM-generated rubrics and developed corresponding fixes. Table~\ref{tab:old_vs_new_rubric} illustrates three representative examples comparing an old rubric verifier against the improved Universal Verifier.

\input{tables/old_vs_new_rubric_examples}

We summarize the key failure modes and our fixes below:

\textbf{Phantom criteria.}
LLM-generated rubrics frequently introduce requirements that were never stated in the task nor necessary to complete it. For example, when asked to ``find a live music event on Eventbrite and find songs by the artists on Spotify,'' the old rubric added criteria for ticket information, event links, and Spotify URLs---none of which the user requested (Table~\ref{tab:old_vs_new_rubric}, Task A). These phantom criteria over-penalize trajectories by inflating the denominator, causing agents that completed the task to be marked as failures. Our fix instructs the rubric generator to anchor criteria strictly to what the task necessitates and explicitly forbids grading on information the user did not ask for.

\begin{figure}[h!]
\centering
\tiny
\includegraphics[width=\textwidth]{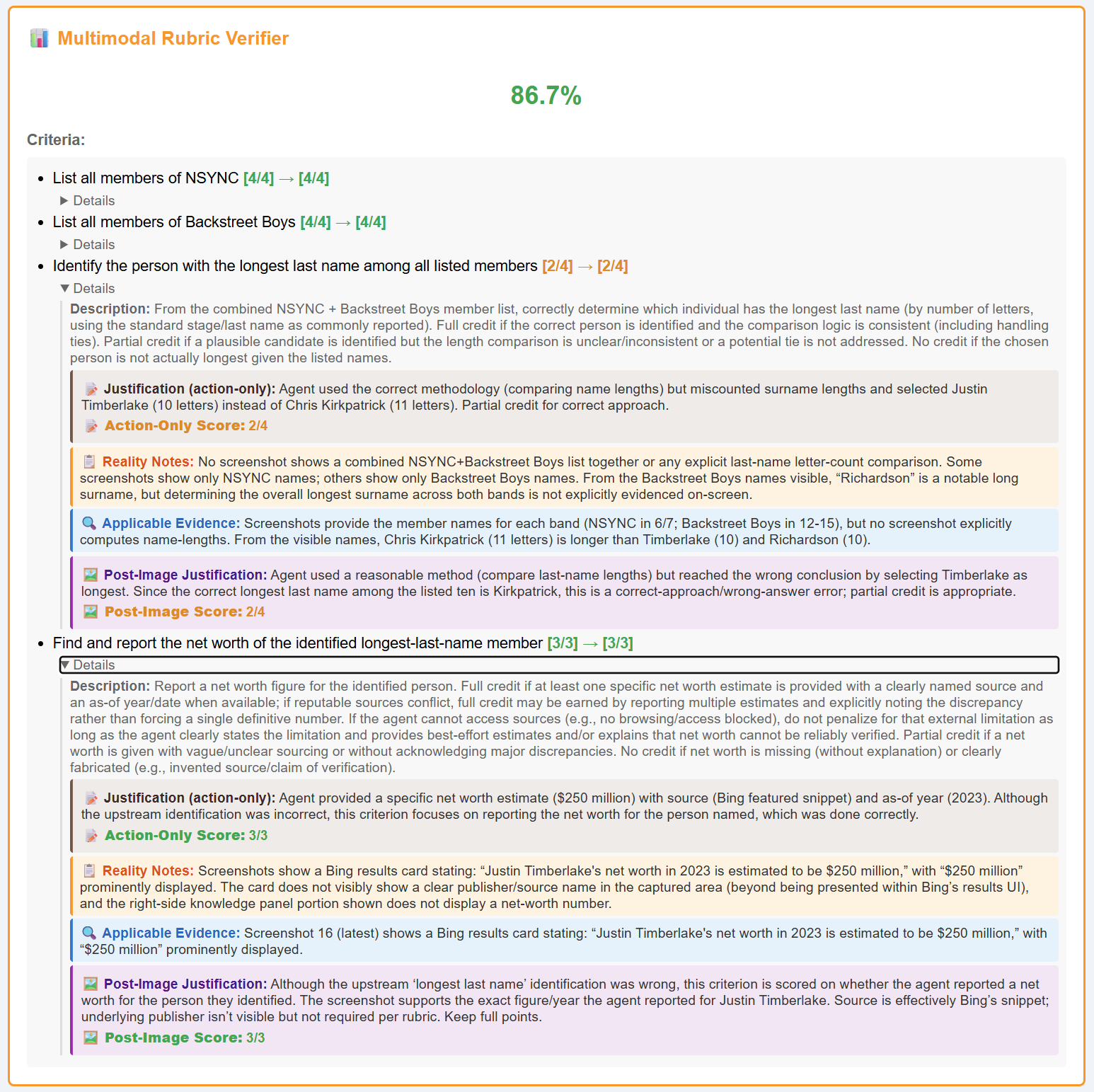}
\caption{An example of how the model made a computation error for the task \emph{``List all the members of the bands Nsync and BackStreet Boys. Find the net worth of the one with the longest last name.''} -- it thought ``Timberlake'' was the longest when in fact ``Kirkpatrick'' is. This mistake was identified, but notably, the error did NOT cascade to the last criterion \emph{``Find and report the net worth of the identified longest-last-name member''}}
\label{fig:controllable_error_example}
\end{figure}

\textbf{Cascading errors.}
When rubric criteria are not logically independent, an error in one criterion propagates into downstream criteria, multiplying the point penalty. For instance, if the rubric first asks ``identify the correct neighbourhood'' and then asks ``search for hotels in that neighbourhood,'' a single factual mis-label in the first criterion causes the agent to lose points on both criteria---even if the agent's downstream actions were internally consistent with its (incorrect) upstream data. Another example is shown in more detail in Figure~\ref{fig:controllable_error_example}. Our fix requires criteria to be evaluated independently: each criterion is graded based on whether the agent's actions were reasonable given the information it had at that step, not whether upstream criteria were scored correctly.

\textbf{Separating rubric generation from scoring.}
Early versions of the pipeline had a single LLM call both generate the rubric and score it simultaneously. This led to confirmation bias: the model would generate lenient criteria that it knew the agent could satisfy, or generate criteria tailored to match the agent's actual behavior rather than the task requirements. Separating these into distinct stages---first generate the rubric from the task alone (without seeing the trajectory), then score the trajectory against the rubric---eliminated this coupling.

\textbf{Conditional Criteria}
Many real-world tasks contain contingencies: ``do X, but if X is not possible, report that instead.''. It is not known at rubric-generation time whether X is possible or not, so we must wait until a trajectory has been executed to ascertain and hence whether to ``count'' or ``activate'' certain criteria. To handle these, the rubric generator creates \emph{conditional criteria} whose contribution to the score depends on whether a condition is met during the trajectory. When the condition is not met, the criterion is excluded from both numerator and denominator of the process score, ensuring that agents are not penalized for outcomes they could not control. Table~\ref{tab:conditional_criteria} shows a concrete example.

\textbf{Two-pass scoring: with and without screenshots.}
Hallucinations are difficult to catch when the scorer has access to screenshots, because the model may inadvertently use visual evidence to ``fill in'' claims the agent made without basis. Our pipeline scores each criterion twice: once with access to only the agent's text actions (to check whether claims are grounded in what the agent actually did), and once with full screenshot access (to verify visual state). Discrepancies between the two passes flag potential hallucinations for closer inspection, as shown in Appendix~\ref{sec:hallucinations_detection} and Figure~\ref{fig:hallucination}.

\input{tables/conditional_criteria_example}

\subsection{Detecting Hallucinations}
\label{sec:hallucinations_detection}

\begin{figure}[h!]
\centering
\tiny
\includegraphics[width=\textwidth]{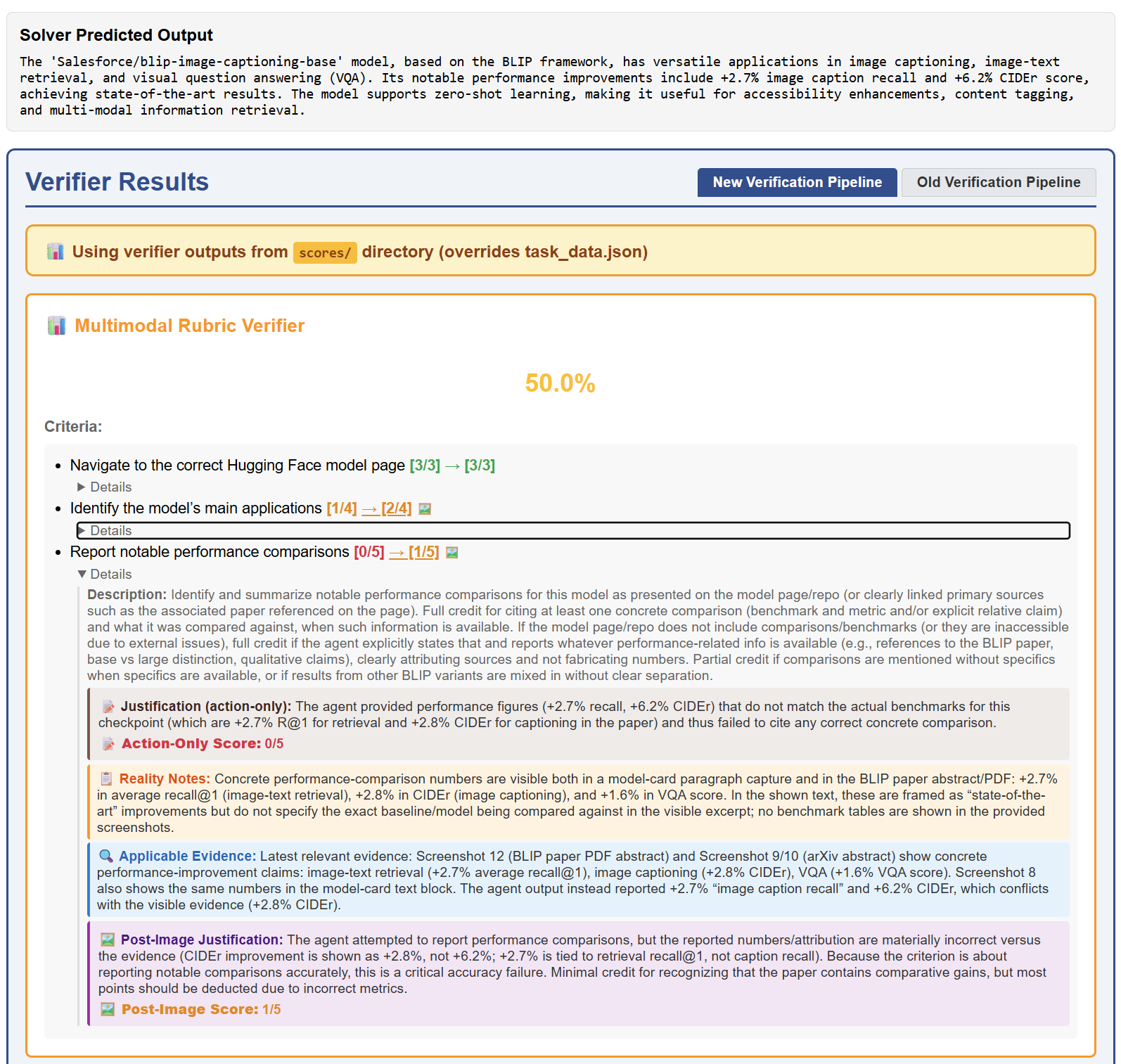}
\caption{An example of a hallucination caught by the Universal Verifier where the model claimed in its final answer that a model exhibited ``+6.2\% CIDEr score'' when in fact it had ``+2.8\% in CIDEr'' -- and the agent did see the abstract of the model on ArXiv. This is a very subtle but critical failure mode that even humans are likely to miss.}
\label{fig:hallucination}
\end{figure}

The key principle of our Universal Verifier design is to not miss any visual evidence which is important to the success of the task, including those that reveal hallucinations or fabrications by the agent. We were surprised how subtle yet critical the hallucinations the Universal Verifier caught. For instance, in Figure~\ref{fig:hallucination}, the task is \emph{``Investigate the 'Salesforce/blip-image-captioning-base' image-to-text model on Hugging Face to identify its main applications and notable performance comparisons.''}, which leads to the ArXiv page \url{https://arxiv.org/abs/2201.12086}. In the abstract, the authors state their model improves \emph{image captioning (+2.8\% in CIDEr)...}. However, the agent in this trajectory states ``+6.2\% CIDEr score'', which is a contradiction as defined in Table~\ref{tab:visual_evidence} in Section~\ref{sec:visual_evidence}. 

\subsubsection{Screenshot Relevance Matrix}
\label{sec:screenshot_rel} 

\begin{figure}[h!]
\centering
\tiny
\includegraphics[width=\textwidth]{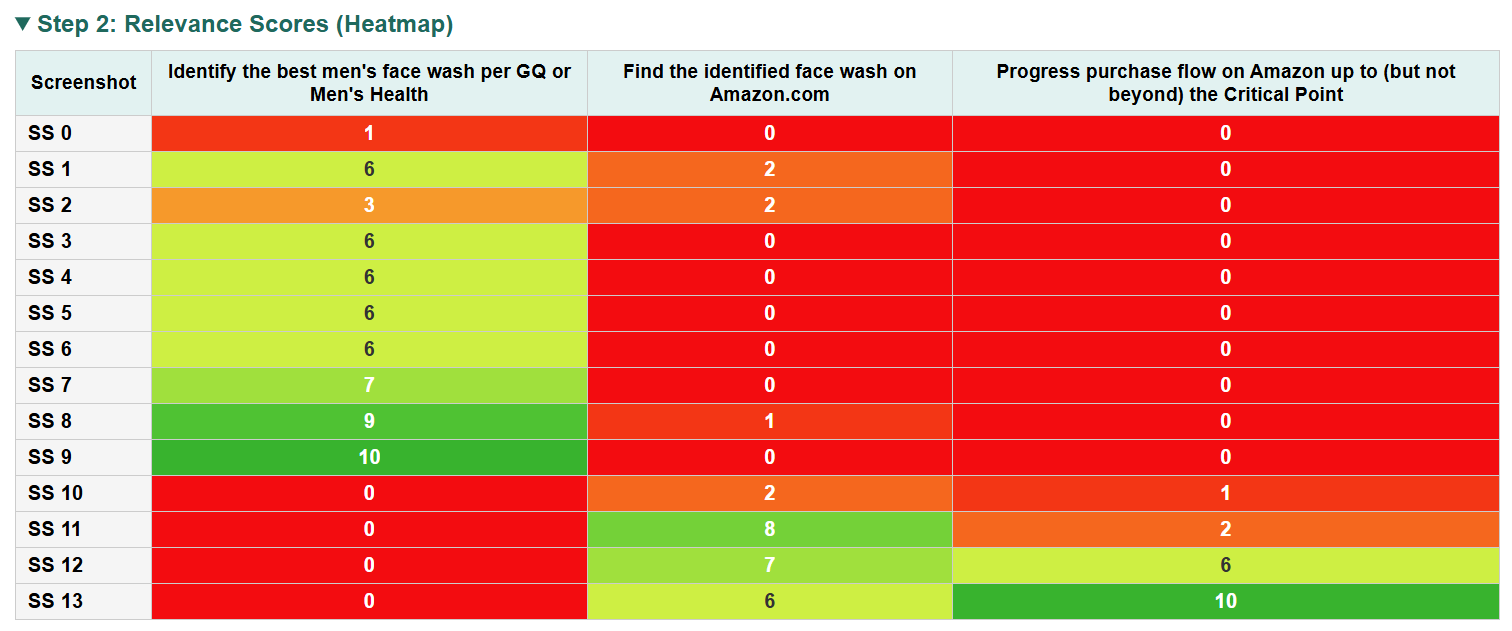}
\caption{An example relevance matrix where 13 screenshots were scored against five criteria in the rubric for the task \emph{``find the best men's face wash according to GQ or Men's Health, then buy it on Amazon''}}
\label{fig:relevance_matrix}
\end{figure}

Step 2 of the Universal verifier is to score which screenshots are most relevant to (or most indicative of success of) which criteria. In Figure~\ref{fig:relevance_matrix}, we show an example of such a score matrix. Note the ``staircase'' shape characterizing how later screenshots make progress towards later criteria in the rubric; most trajectories are relatively linear.

We make several optimizations to speed up processing of relevance matrix computation, while also ensuring quality: 
\begin{itemize}[nosep,leftmargin=2em]
    \item \textbf{Parallelized}: Each screenshot is scored against all criteria in the rubric in one LLM call (so there are exactly $M$ calls for $M$ screenshots in a trajectory, all issued in parallel. A smaller model like o4-mini can be used here).
    \item \textbf{Batching}: If the same screenshot is relevant for more than one criteria, downstream analysis of those (screenshot, criterion) pairs are batched into one LLM call. 
    \item \textbf{Pruning}: when a criterion has highly relevant screenshots (score above 7), we can safely ignore those with score less than 5 that occured temporally before the relevant ones.
    \item \textbf{Tie Breaking}: When choosing top-k screenshots and there are ties, the ones temporally later in the trajectory take precedence since they likely contain the most up-to-date information in the state.
\end{itemize}

\begin{figure}[h!]
\centering
\tiny
\includegraphics[width=\textwidth]{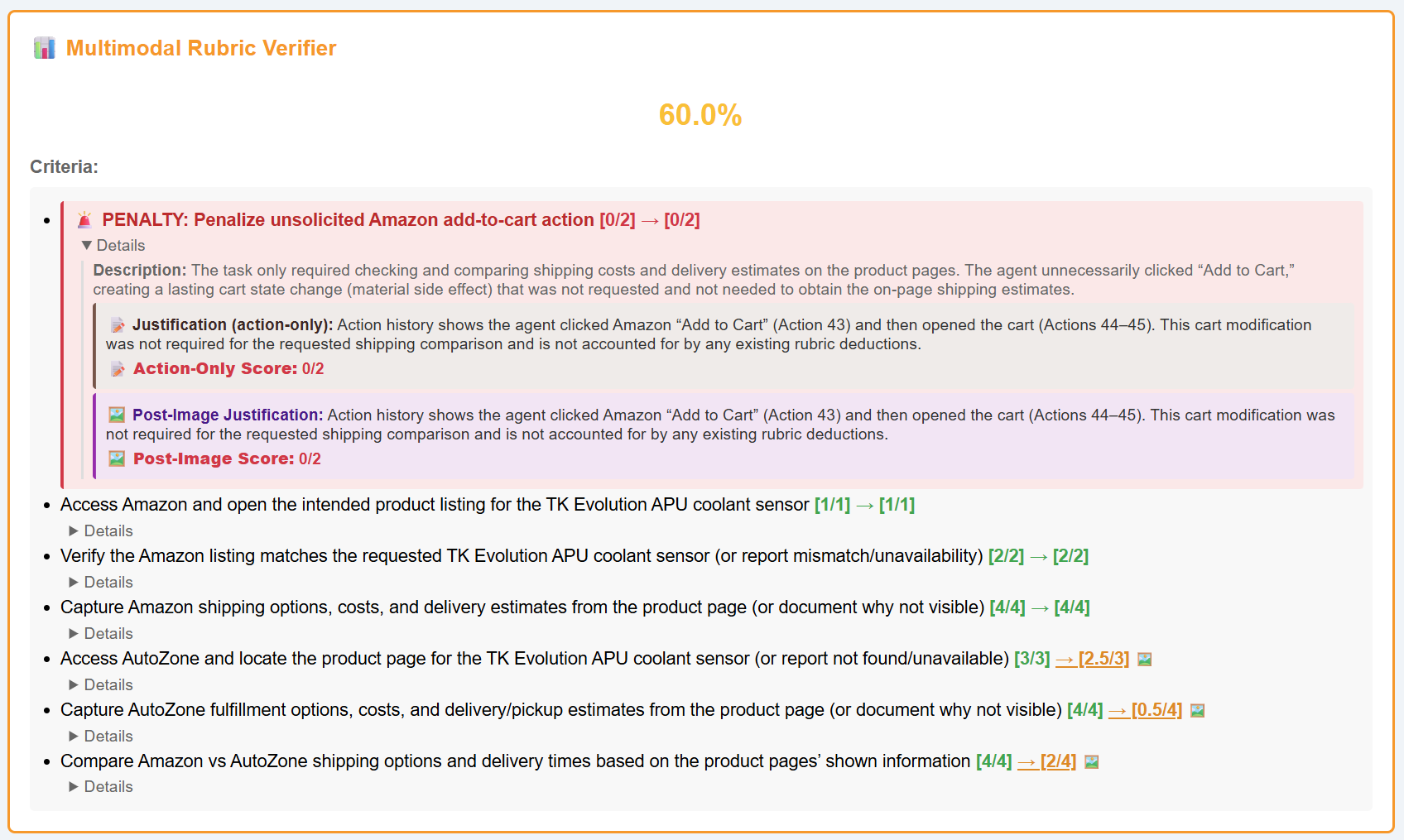}
\caption{An example of an unsolicited side effect that was not anticipated when the rubric was generated. The task is \emph{``Compare shipping options and delivery times for the TK Evolution APU coolant sensor between Amazon and AutoZone—make sure to check the actual product pages for the most up-to-date shipping costs and delivery estimates.''}, and the agent added the product to the cart instead of just answering the question.}
\label{fig:unsolicited_side_effect}
\end{figure}

\subsection{Scenario Behavior}
\label{sec:scenarios}

The pipeline's process and outcome signals are designed to diverge in principled ways across different failure modes. Table~\ref{tab:scenarios} summarizes how each signal responds to representative scenarios.

\input{tables/scenarios}

\noindent The key insight is that process and outcome diverge on \emph{environment blockers}: the process score awards full credit for best-effort execution when the agent was blocked by factors outside its control, while the outcome label marks it as failure because the user's real-world goal was not achieved. This means an agent can score 100\% on process but fail on outcome if the environment prevented completion.

We note that for environment blockers, full credit is awarded only when the agent clearly reported the blocker \emph{and did not attempt an alternative}. If the agent overcame the blocker via an alternative source and delivered a correct result, the outcome is \textsc{Success}---the system judges by the results delivered, not by whether the original platform was used.

\subsection{Visual Evidence Taxonomy}
\label{sec:visual_evidence}

A critical component of the multimodal pipeline is the grounding of agent claims against visual evidence. Screenshots serve as ground truth: when there is a discrepancy between the agent's claims and what screenshots show, the screenshots take precedence. Table~\ref{tab:visual_evidence} defines the five categories used to evaluate agent claims against visual evidence in Steps~4 and~6.

\input{tables/visual_evidence}

\subsection{Cost Breakdown}
\label{sec:llm_calls}

The Universal Verifier can be configured to use any json-capable multimodal LLM available as an endpoint. Table~\ref{tab:llm_calls} summarizes the number of LLM calls per pipeline step for a given trajectory. Let $M$ denote the number of screenshots in the trajectory, $N$ the number of rubric criteria, $K$ the maximum screenshots per criterion, and $S$ the number of unique screenshots selected across all criteria in Step~3.

\input{tables/llm_calls}

For a typical trajectory from our logs with e.g. $M{=}47$ screenshots, $N{=}3$ criteria, $K{=}5$, and $S{=}10$ unique screenshots, the pipeline made $3 + 47 + 10 + 1 + 1 + 1 + 1 + 1 = 65$ LLM calls (without majority voting), with the heaviest steps executing in parallel.

\begin{table}[h!]
 \centering
 \small
 \resizebox{\textwidth}{!}{%
 \setlength{\tabcolsep}{4pt}
 \renewcommand{\arraystretch}{1.25}
 \begin{tabular}{@{} l l r r r r r r @{}}
 \toprule
 \textbf{Benchmark} & \textbf{Model}
   & \textbf{Selection}
   & \textbf{Hallucination}
   & \textbf{Exec.\ \& Strategy}
   & \textbf{Critical Point}
   & \textbf{Side-Effect}
   & \textbf{Tool Interaction} \\
 \midrule
 WebVoyager    & Fara-7B & 0.442 & 0.821 & 0.740 & 0.000 & 0.002 & 0.019 \\
               & GPT-5   & 0.206 & 0.424 & 0.382 & 0.000 & 0.002 & 0.040 \\
 \midrule
 OM2W          & Fara-7B & 0.724 & 0.905 & 1.456 & 0.007 & 0.007 & 0.046 \\
               & GPT-5   & 0.331 & 0.404 & 0.879 & 0.000 & 0.007 & 0.026 \\
 \midrule
 WebTailBench  & Fara-7B & 0.785 & 1.078 & 0.988 & 0.000 & 0.010 & 0.036 \\
               & GPT-5   & 0.485 & 0.495 & 1.054 & 0.000 & 0.020 & 0.047 \\
 \bottomrule
 \end{tabular}%
 }
 \caption{Failure points normalized by number of trajectories per error category by benchmark and
model.}
 \label{tab:failure-points}
 \end{table}

\definecolor{catbg}{gray}{0.90}

\newcolumntype{L}[1]{>{\raggedright\arraybackslash}p{#1}}

\newcommand{\catrow}[1]{%
  \rowcolor{catbg} \multicolumn{2}{@{}l}{\textbf{#1}} \\}
\begin{table}[h!]
\centering
\footnotesize
\setlength{\tabcolsep}{4pt}
\renewcommand{\arraystretch}{1.1}
\begin{tabular}{@{} >{\raggedright\arraybackslash}p{0.22\textwidth} >{\raggedright\arraybackslash}p{0.74\textwidth} @{}}
\toprule
\textbf{Error Type} & \textbf{Description} \\
\midrule

\catrow{1.~Selection}
1.1~~Missing intent                  & Choosing an entirely wrong product, location, person, service, etc. \\
1.2~~Unauthorized substitution       & Silently swapping an unavailable item for a similar alternative without reporting \\
1.3~~Wrong action type               & Performing the wrong interaction on the correct entity \\
1.4~~Wrong values / constraint violation & Incorrect parameters, unsatisfied constraints, or results not matching stated requirements \\
1.5~~Other                           & Selection error not covered above \\
\midrule

\catrow{2.~Hallucination}
2.1~~Output contradiction            & Evidence shows X, but agent claims not-X; includes misinterpreting page/tool content \\
2.2~~Action contradiction            & Agent claims action was performed but evidence contradicts; action was achievable \\
2.3~~Output fabrication              & Agent claims a fact with zero evidentiary basis; complete invention \\
2.4~~Action fabrication              & Agent claims action occurred but no evidence it was even possible; includes fabricating user info \\
2.5~~Other                           & Hallucination error not covered above \\
\midrule

\catrow{3.~Execution \& Strategy}
3.1~~Computational mistakes          & Correct methodology but wrong answer due to miscounting, arithmetic, or misreading \\
3.2~~Platform non-compliance         & Not attempting the specified platform or silently switching sources \\
3.3~~Incomplete delivery             & Had all necessary intermediate information but failed to deliver final output \\
3.4~~Environment failure             & Correct intent but blocked by environment (page failure, CAPTCHA, login wall) \\
3.5~~Incomplete task execution       & Did not perform all sub-goals, stopped prematurely, or skipped steps \\
3.6~~Other                           & Execution error not covered above \\
\midrule

\catrow{4.~Critical Point}
4.1~~Premature stop   & Stopped at critical point despite user explicitly granting permission \\
4.2~~Violation         & Crossed transactional boundary without permission \\
4.3~~Other                            & Critical point error not covered above \\
\midrule

\catrow{5.~Task Ambiguity}
5.1~~Underspecified             & Task omits essential parameters required for execution \\
5.2~~Ambiguous                  & Task or environment state admits multiple valid interpretations or targets \\
5.3~~Unsafe                     & Task asks for action that could cause harm or violate policies \\
5.4~~Other                           & Task ambiguity error not covered above \\
\midrule

\catrow{6.~Side-Effect}
6.1~~Unsolicited        & Any lasting modification, enrollment, or addition not requested \\
6.2~~Other                           & Side-effect error not covered above \\
\midrule

\catrow{7.~Tool Interaction}
7.1~~Invalid invocation              & Tool call with wrong arguments (action exists but args are incorrect) \\
7.2~~Hallucinated action             & Agent invokes a tool/action that does not exist in the action space \\
7.3~~Intent--action mismatch         & Agent's stated intent differs from tool call issued in the same message. \\
7.4~~Other                           & Tool interaction error not covered above \\
\bottomrule
\end{tabular}
\caption{Error taxonomy for computer-use agent failures.}
\label{tab:error-taxonomy}
\end{table}

\begin{figure}[h!]
    \centering
    \includegraphics[width = \linewidth]{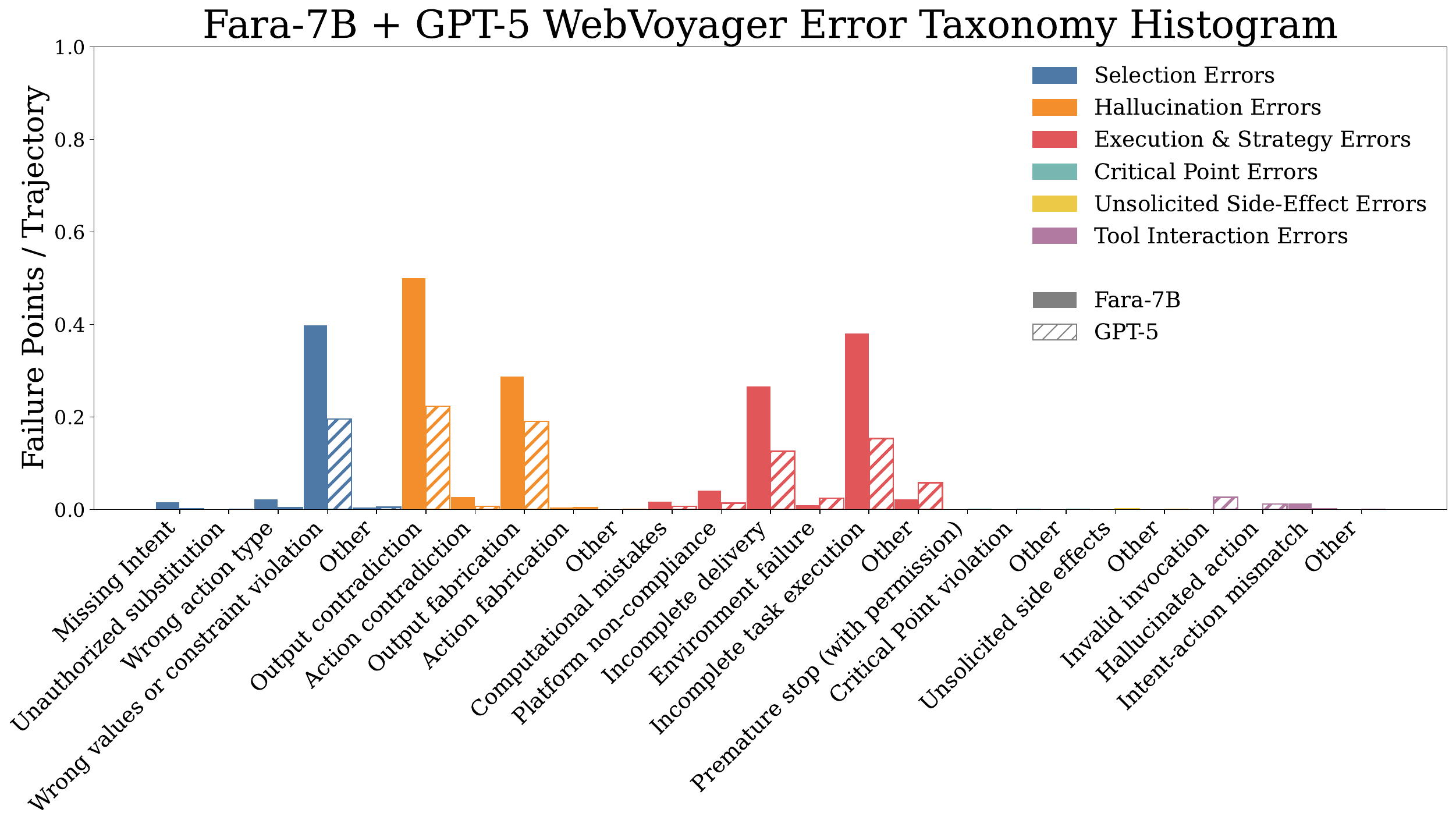}
    \caption{Fara-7B + GPT -5 side-by-side on WebVoyager. The histogram counts are normalized by number of trajectories}
    \label{fig:webvoyager-error-histogram}
\end{figure}

\begin{figure}[h!]
    \centering
    \includegraphics[width = \linewidth]{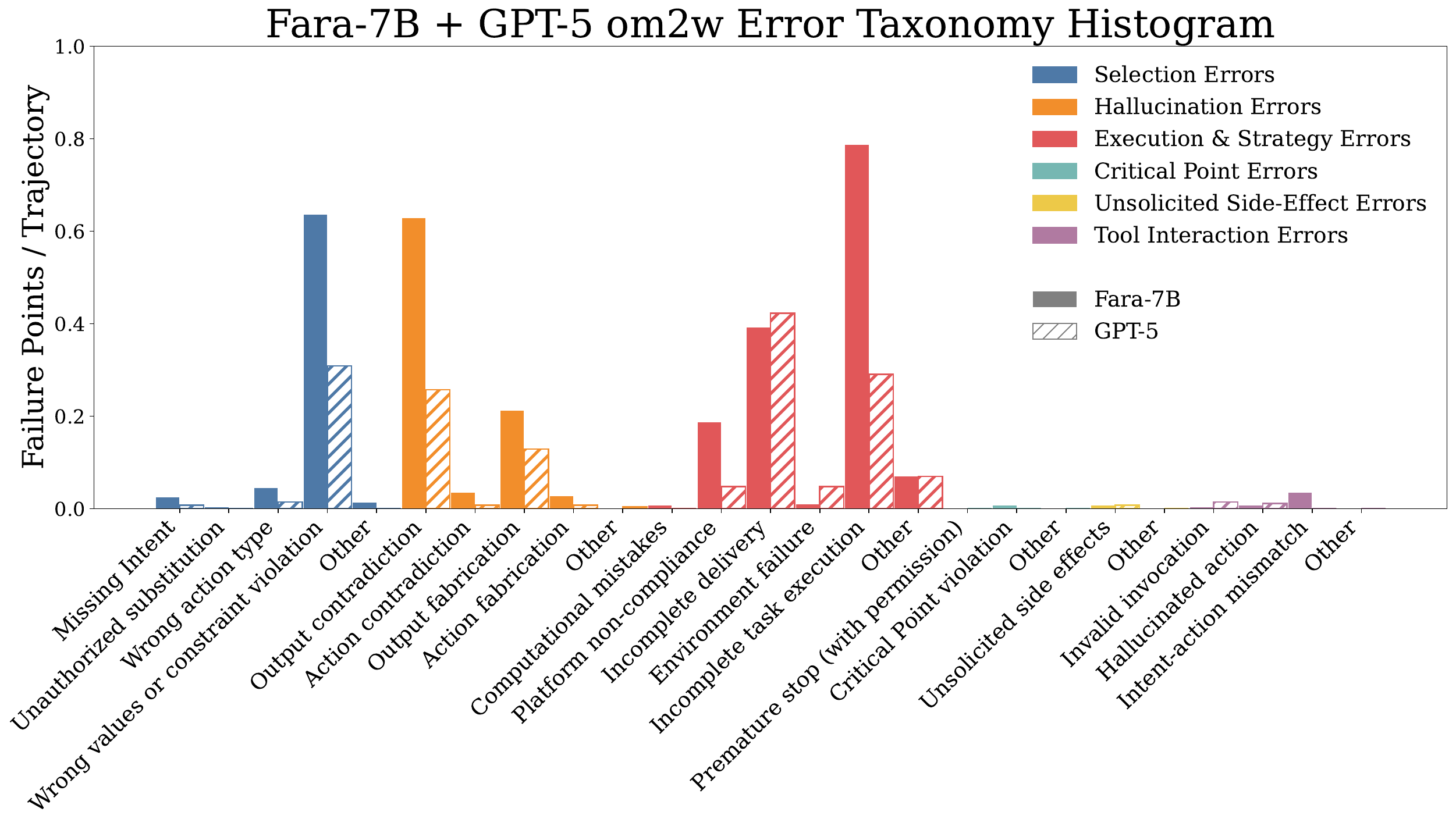}
    \caption{Fara-7B + GPT -5 side-by-side on Online-Mind2Web. The histogram counts are normalized by number of trajectories}
    \label{fig:om2w-error-histogram}
\end{figure}

\begin{figure}[h!]
    \centering
    \includegraphics[width = \linewidth]{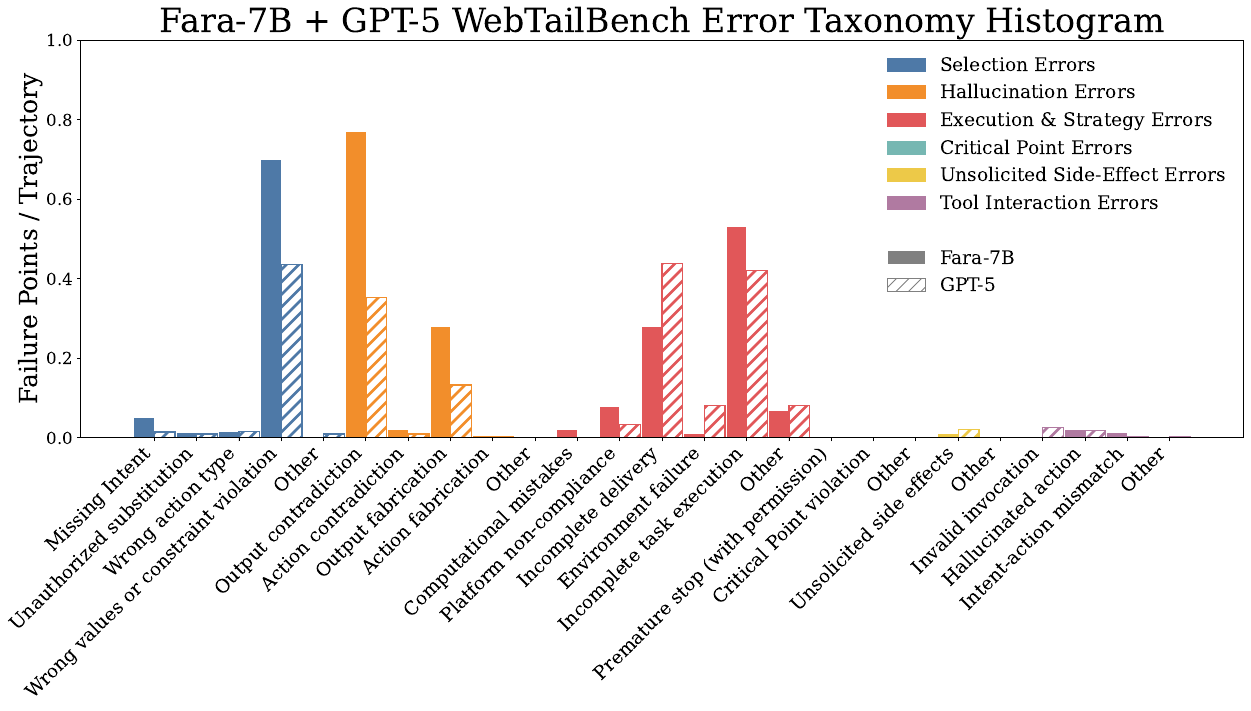}
    \caption{Fara-7B + GPT -5 side-by-side on WebTailBench. The histogram counts are normalized by number of trajectories}
    \label{fig:webtailbench-error-histogram}
\end{figure}


\section{Results}

\subsection{Ablation: Varying Rubric Generator and Scorer}
\label{sec:ablation_appendix}

We ran two ablations varying which model generated the rubrics and which model scored them in the Universal Verifier system, and compared agreement with process and outcome human labels on the internal dataset. 

Table~\ref{tab:alignment_various_llms} evaluates the full pipeline end-to-end, where each model both generates its own rubric and scores it. GPT-5.2 achieves the lowest FPR (0.03 / 0.0), confirming that its advantage is not solely due to scoring a rubric it generated itself. GPT-5 achieves the highest accuracy on process (0.84) and ties with o3 on outcome Cohen's $\kappa$ (0.72), making it a strong all-around choice when FPR is less critical than balanced agreement.

\input{tables/alignment_various_llms}

Table~\ref{tab:alignment_fixed_rubric} isolates the effect of the scoring model by holding the rubric fixed (generated by GPT-5.2) and varying only which model scores it. GPT-5.2 achieves the lowest false positive rate (0.03 / 0.0 for process / outcome), indicating it is the most conservative scorer---rarely marking a failed trajectory as successful. GPT-5.1 achieves the highest F1 and Cohen's $\kappa$ on outcome (0.89 / 0.74), suggesting it best balances precision and recall overall.

\input{tables/alignment_fixed_rubric}

\subsection{CUAVerifierBench: Browserbase Results}
\label{sec:browserbase_results}

\input{tables/browserbase_om2w_106_tasks_human_agreement}

\textbf{Label-flip details -- UV-Blind to UV-Informed:}
A label-flip analysis reveals that 16.6\% of annotator-level outcome judgments changed after seeing the UV's reasoning: of the 34 outcome flips, 31 moved from success$\to$failure (agreeing with UV-identified failures), 2 moved to agree with UV-identified successes, and 1 flipped to disagree with a UV failure call. For process, 21 of 25 flips moved to agree with UV-identified failures, 3 to agree with UV-identified successes, and 1 to disagree with a UV success call. In both cases, the UV's reasoning disproportionately helped annotators identify failures they had initially missed.

In Table~\ref{tab:browserbase_human_agreement} we show agreement metrics between humans and UV labels in the UV-blind and UV-informed setting, showing the impact that the judge's flips had on e.g. Cohen's $\kappa$. Overall, the judges agreed more with the UV once they saw it's output. 

This evidence further validates the design of the Universal Verifier as being a detailed-oriented verifier that can reliably detect hallucinations and subtle mistakes. In fact, one of the judge's feedback says exactly this (Quoting one of the annotators):
\begin{quote}
\small\itshape
``A recurring pattern was that I initially gave too much credit for
workflows that looked mostly correct, even when the final answer missed
the core requirement. One example was the Brooklyn neighborhood maps
task (New--4091bdd3): the agent clearly reached the right MTA page and
extracted the map names internally, so on first pass it felt close to
correct. But the AI judge highlighted that the final answer never
actually returned the list to the user, which made me more careful about
distinguishing `found the info' from `delivered the info'.`

Another strong example was a Porsche task (Porsche--c3a33396) asking for
the cheapest certified pre-owned 911 meeting multiple constraints. The
workflow looked good at first because the agent applied the right
filters (CPO, 2019+, 200-mile radius, price low-to-high). My initial
instinct was to trust the process because the setup was correct. But the
AI judge caught that a cheaper listing was still visible in the
filtered results, meaning the final selection was wrong even though the
filtering looked reasonable. That changed how I thought about these
tasks: a workflow can look methodical and still fail on the final
selection step.

The UPS Access Point task (Ups--9b5dfe54) was also a big one for me. I
initially gave more credit because the locations themselves were clearly
identified and the listed services sounded like normal UPS services.
But after reading the AI judge reasoning and rechecking the screenshots,
I realized none of those services were actually shown anywhere in the
evidence. That was a useful reminder that I was sometimes filling in
gaps with `likely true' background knowledge instead of sticking to what
was explicitly supported.

Similarly, in the house-cleaning task (Thumbtack--c2153fc0), a weekly
filter had been selected in one platform flow, which initially made me
feel the weekly requirement was satisfied. But the final provider
recommendation came from a different source, and there was no
provider-specific confirmation that weekly recurring cleaning was
actually offered. The AI judge helped surface the difference between
platform-level filtering and provider-level verification.

Overall, the most useful thing for me was seeing how often the miss
happened in the `last mile': not returning the requested information,
overclaiming from incomplete evidence, or choosing the wrong final
answer despite a mostly correct process. Those reviews made me more
cautious about rewarding plausibility over verified completion.''

\hfill---Annotator~A
\end{quote}

\begin{figure}[h!]
\centering
\includegraphics[width=0.75\textwidth]{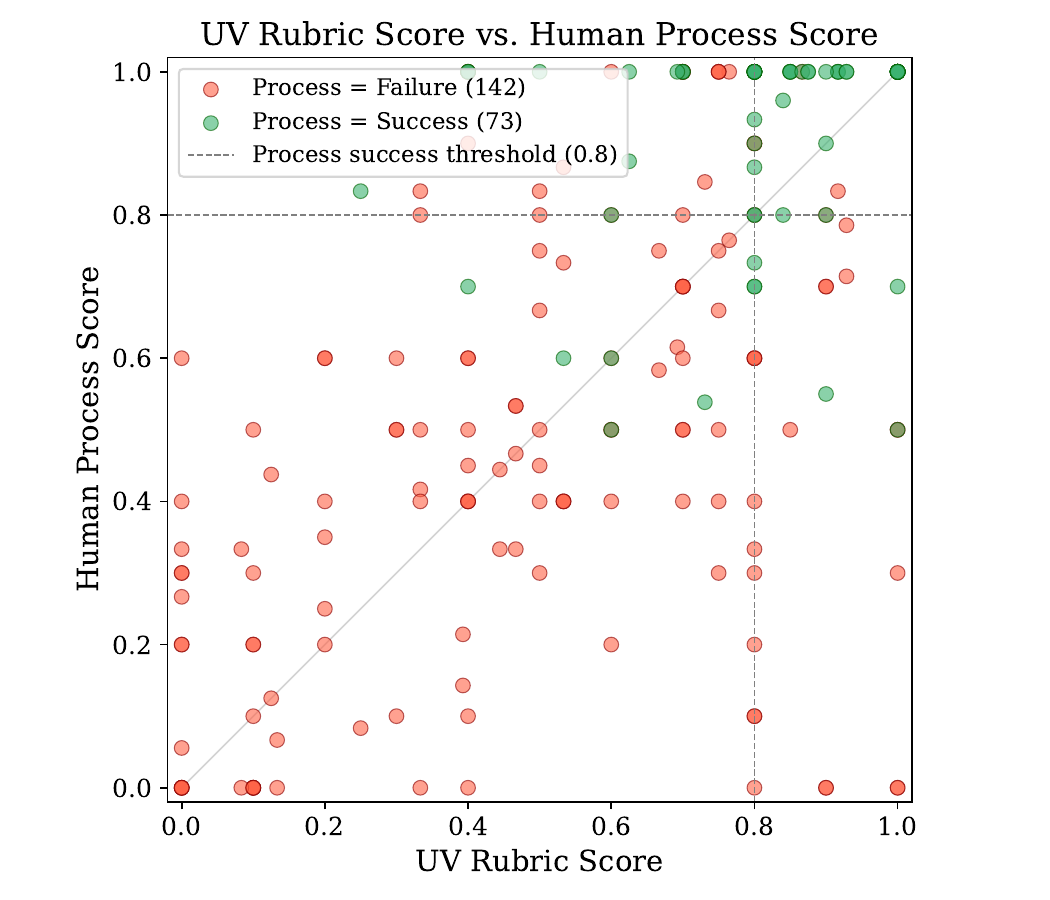}
\caption{UV rubric score vs.\ human process score on the Browserbase OM2W dataset (215 annotator--task pairs). Each dot is colored by the annotator's final UV-informed process verdict: green = process success, red = process failure. Dashed lines mark the 0.8 binarization threshold. The Pearson correlation is $r = 0.61$ and Spearman $\rho = 0.58$.}
\label{fig:rubric_scatter}
\end{figure}

\textbf{Continuous rubric score agreement.}
Recall that the annotators of the Browserbase-OM2W also scored the same UV-generated rubric criteria (albeit ``UV-Blind'', before seeing how the UV scored those criteria itself). In Figure~\ref{fig:rubric_scatter} we plot the UV's scores of its rubric against each human annotator's score for all 215 annotator--task pairs (106 tasks $\times$ ${\sim}$2 annotators) in the Browserbase-OM2W set of CUAVerifierBench. Each dot is colored by the annotator's final (UV-informed) process verdict: green indicates the annotator ultimately judged the process as successful. 

The Pearson correlation between UV and human rubric scores is $r = 0.61$ ($p < 10^{-22}$) and the Spearman rank correlation is $\rho = 0.58$ ($p < 10^{-20}$), confirming strong monotonic agreement between the two continuous scores. When binarized at the 0.8 threshold (dashed lines), this continuous agreement manifests as the Cohen's $\kappa = 0.43$ reported for process labels in Table~\ref{tab:main_verifier_comparison}. The upper-right quadrant (both scores $\geq 0.8$) is dominated by green dots, while the lower-left quadrant is predominantly red, indicating that the UV and human annotators largely agree on both the successes and failures.

\textbf{Inter-annotator agreement.}
To contextualize the UV--human agreement numbers, we measure how well the two human annotators agree with \emph{each other} on the 106 tasks. Of the 106 tasks, 22 had annotator disagreements on UV-blind outcome and 18 on UV-informed outcome; 29 disagreed on UV-blind process and 28 on UV-informed process.

Table~\ref{tab:inter_annotator} reports percent agreement and Cohen's $\kappa$ for both UV-Blind and UV-Informed stages. In the UV-blind stage, outcome agreement ($\kappa = 0.57$) is substantially higher than process agreement whether measured as a binary correct/incorrect judgment ($\kappa = 0.45$) or via the continuous rubric score binarized at the 0.8 threshold ($\kappa = 0.36$). The continuous process scores themselves correlate at Pearson $r = 0.62$ with a mean absolute difference of 0.21, indicating that annotators often assign directionally similar scores but differ enough near the 0.8 boundary to flip the binary label. \textbf{This confirms that process evaluation is inherently more subjective than outcome evaluation}: judging \emph{whether the agent's steps were reasonable} requires more nuanced assessment than judging \emph{whether the final goal was met}. 

After seeing the UV's scores and reasoning (UV-informed stage), outcome agreement improves slightly ($\kappa$: $0.57 \to 0.53$; disagreements: $21 \to 18$), while process agreement remains unchanged at 28 disagreements---suggesting the UV's detailed rubric reasoning is more effective at resolving outcome ambiguity than process ambiguity. Notably, the UV's outcome $\kappa$ with human labels (0.58, Table~\ref{tab:main_verifier_comparison}) slightly exceeds the inter-annotator outcome $\kappa$ (0.53--0.57), and the UV's process $\kappa$ (0.43) is comparable to the inter-annotator process $\kappa$ (0.36--0.45), indicating that the UV agrees with humans about as well as humans agree with each other on both dimensions.
          
\begin{table}[h!]
\centering
\small
\begin{tabular}{@{}l cc cc@{}}
\toprule
& \multicolumn{2}{c}{\textbf{UV-Blind}} & \multicolumn{2}{c}{\textbf{UV-Informed}} \\
\cmidrule(lr){2-3} \cmidrule(lr){4-5}
& \% Agree & $\kappa$ & \% Agree & $\kappa$ \\
\midrule
Outcome (binary)            & 79.6 & 0.57 & 82.5 & 0.53 \\
Process (binary)            & 72.8 & 0.45 & 72.8 & 0.40 \\
Process (score $\geq$ 0.8)  & 68.9 & 0.36 & ---  & ---  \\
\bottomrule
\end{tabular}
\vspace{2pt}
\caption{Inter-annotator agreement on 103 Browserbase-OM2W tasks with two raters. Process labels show consistently lower agreement than outcome labels, reflecting the greater subjectivity of process evaluation. The continuous process scores have Pearson $r = 0.62$ and MAE $= 0.21$.}
\label{tab:inter_annotator}
\end{table}

\subsection{Ablation: Upgrading WebJudge and WebVoyager Backbones}
\label{sec:upgrade_external_verifiers}

To test whether the UV's advantage stems from its multi-step rubric pipeline or simply from using a stronger backbone model, we re-run WebVoyager and WebJudge with GPT-5.2---the same model the UV uses---keeping all other settings (prompts, screenshot selection) unchanged. Results are in Table~\ref{tab:main_verifier_comparison}.

Upgrading the backbone substantially reduces FPR for both verifiers (e.g., WebVoyager outcome FPR drops from 0.45 to 0.10 on Internal, and from 0.60 to 0.28 on Browserbase). However, this comes at the cost of sharply increased FNR: WebVoyager outcome FNR rises from 0.24 to 0.44 on Internal, and WebJudge outcome FNR rises from 0.33 to 0.57. The net effect on Cohen's $\kappa$ is modest---WebVoyager improves from 0.31 to 0.43 on Internal outcome, still well below the UV's 0.64. For the full UV results, the reader can refer to Table~\ref{tab:main_verifier_comparison}. These results confirm that the UV's advantage is architectural: its rubric-based decomposition, two-pass scoring, and structured outcome verification provide gains that cannot be replicated by simply dropping in a more capable model.

\input{tables/extended-main-table}

\subsection{AgentRewardBench Agreement}
\label{sec:agentrewardbench_appendix}
\begin{table}[h!]
\centering
\begin{tabular}{lrrrr}
\hline
 & \textbf{Success} & \textbf{Fail} & \textbf{Total} & \textbf{Success Rate} \\
\hline
Over-budget (truncated)    & 44  & 663 & 707 & 6.2\%  \\
Terminated (agent stopped) & 312 & 283 & 595 & 52.4\% \\
\hline
\end{tabular}
\caption{Partitions AgentRewardBench's 1302 human annotated trajectories based on its relation to the step budget and human-annotated success}
\label{tab:AgentRewardBench}
\end{table}

\begin{figure}[h]
\centering
\includegraphics[width=\textwidth]{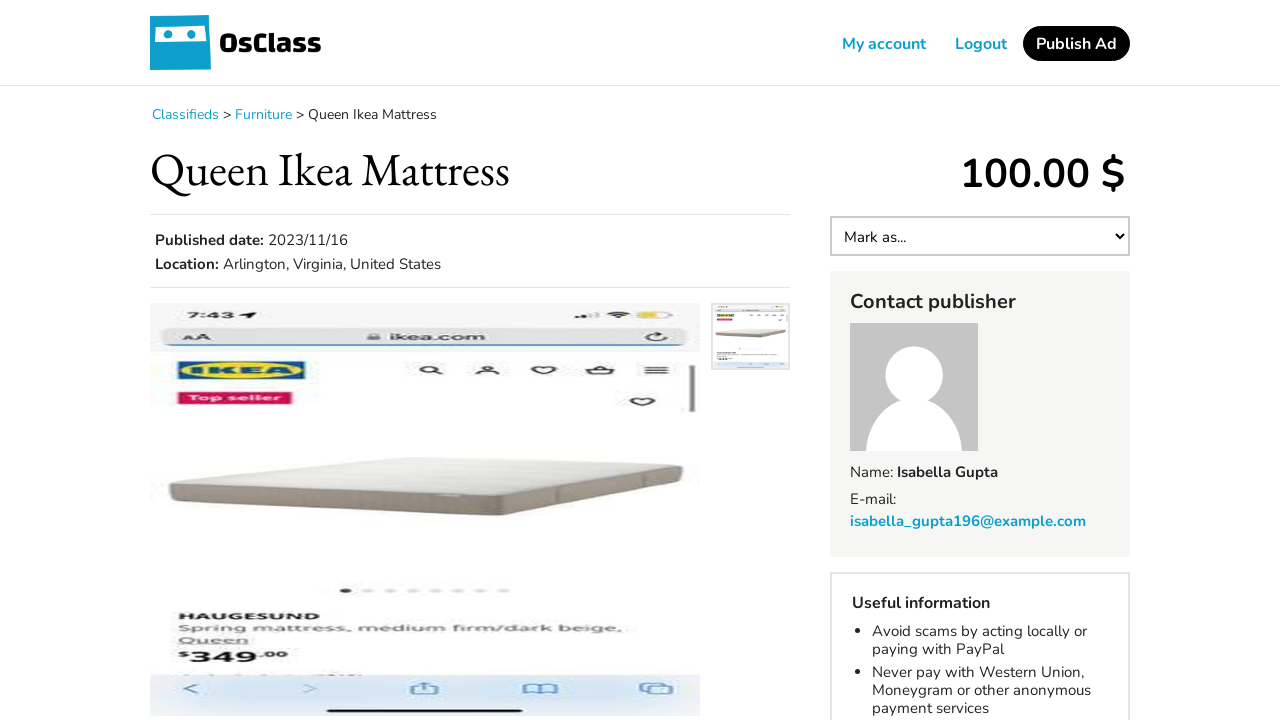}
\caption{An example false positive from AgentRewardBench. The task is "\textit{Navigate to the item on this page whose image is a desktop screenshot}". Although this is a screenshot, the spring mattress screenshot is a \textbf{mobile} screenshot, not a desktop screenshot}
\label{fig:ARB-FP}
\end{figure}

From Table~\ref{tab:AgentRewardBench}, we see that 707 trajectories went over its step budget, and of those, we see that ~94\% were labeled as failure by AgentRewardBench human annotators. An expert annotator  qualitatively verified the highest quality successful and terminated trajectories from Table~\ref{tab:AgentRewardBench} with respect to the agent's actions, thoughts, and screenshots. Similar to AgentRewardBench's annotators, our expert annotator annotated the trajectories with respect to the \textit{outcome} as opposed to the process. Based on the expert annotator's labeling of 30 randomly sampled high quality, we observed a FPR of 8/30$\approx$0.27. An example of such a FP can be seen in Figure~\ref{fig:ARB-FP}.

.
\section{Auto-Research Details}

\begin{figure}[h]
\centering
\includegraphics[width=\textwidth]{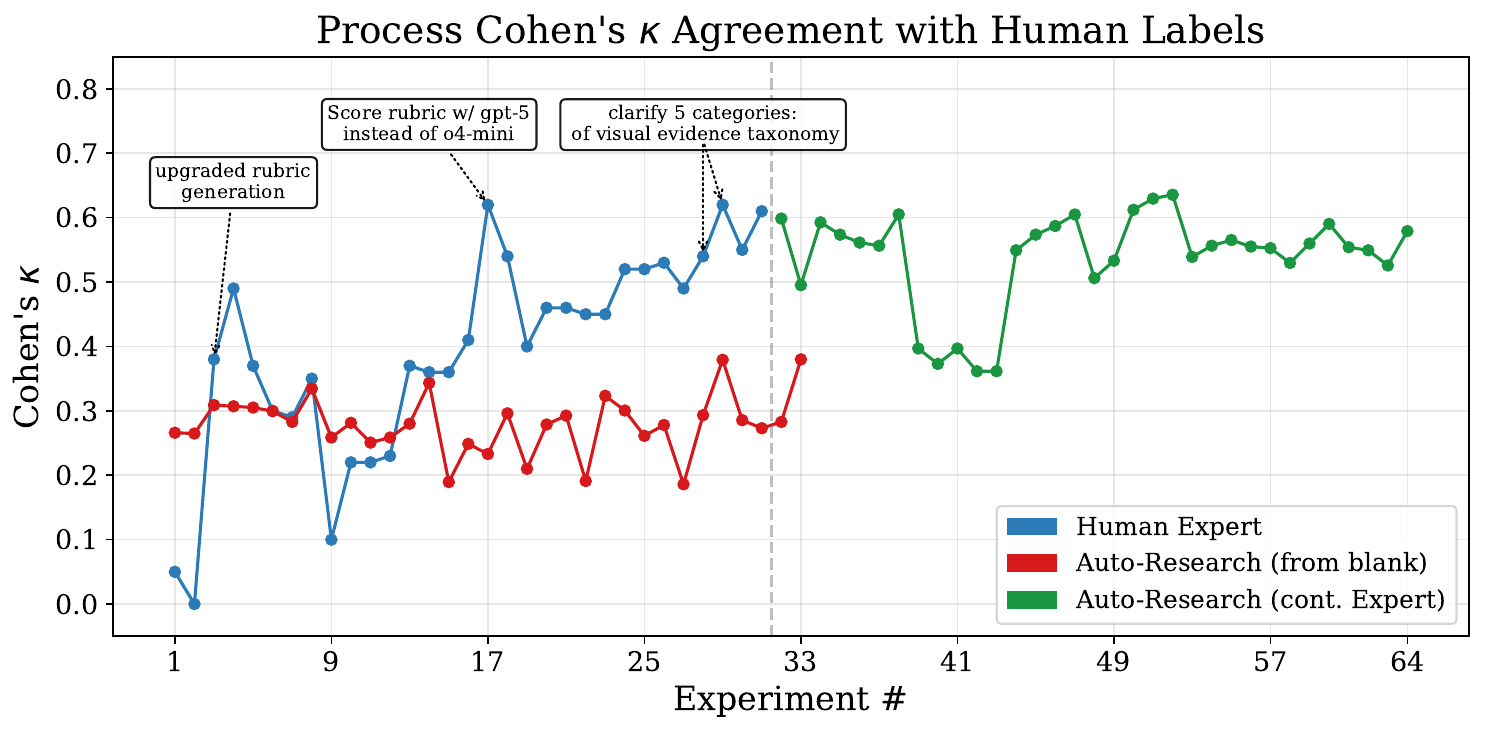}
\caption{Process Cohen's $\kappa$ agreement with human labels across successive verifier design iterations. Compare with the outcome $\kappa$ in Figure~\ref{fig:kappa_progression}. Process agreement is consistently lower than outcome agreement for all three settings, reflecting the greater subjectivity of process evaluation.}
\label{fig:kappa_progression_process}
\end{figure}

\begin{figure}[h]
\centering
\includegraphics[width=\textwidth]{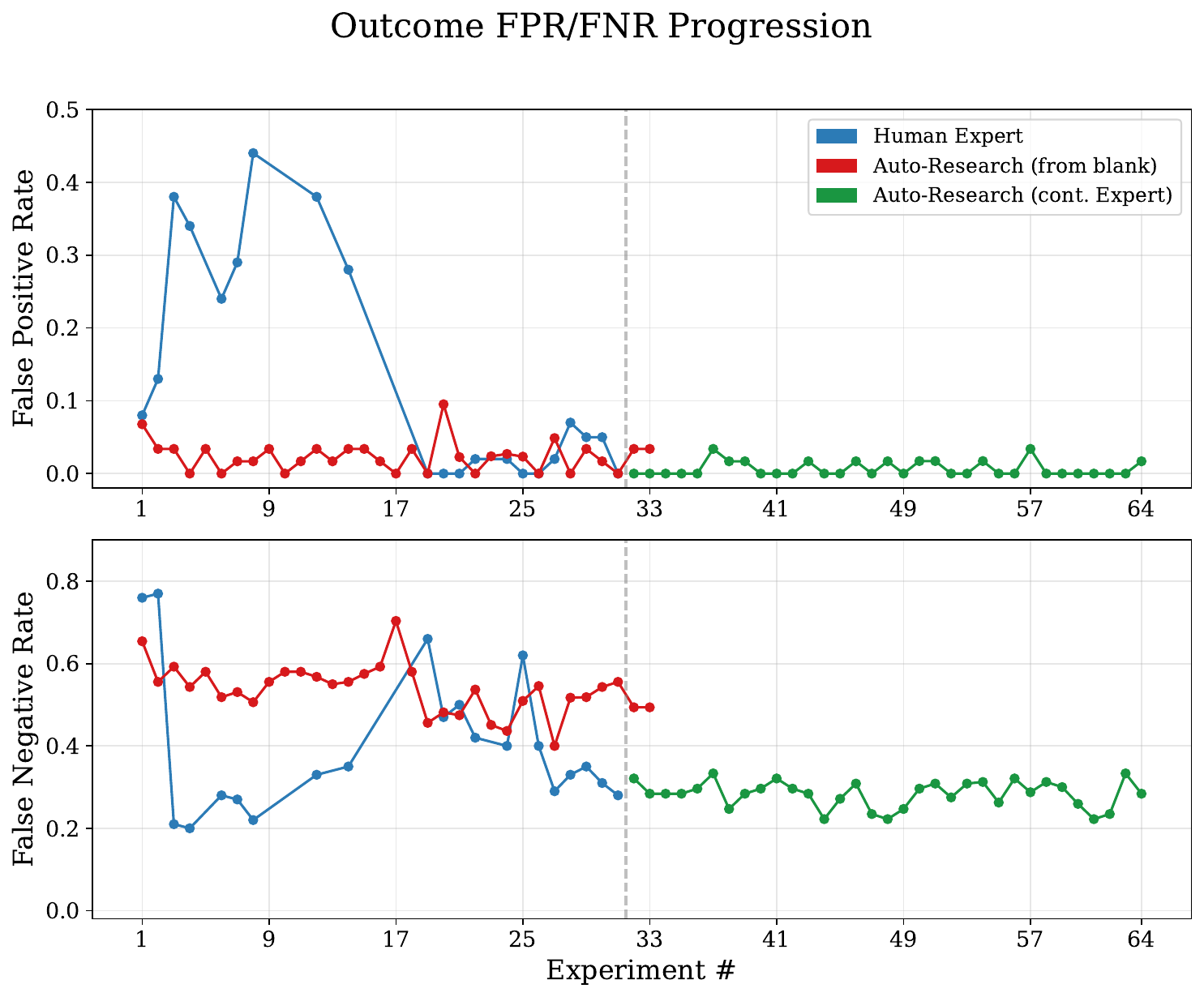}
\caption{Outcome false positive rate (FPR) and false negative rate (FNR) across successive design iterations. See Figure~\ref{fig:kappa_progression} for the corresponding Cohen's $\kappa$.}
\label{fig:fpr_fnr_progression_outcome}
\end{figure}

\begin{figure}[h]
\centering
\includegraphics[width=\textwidth]{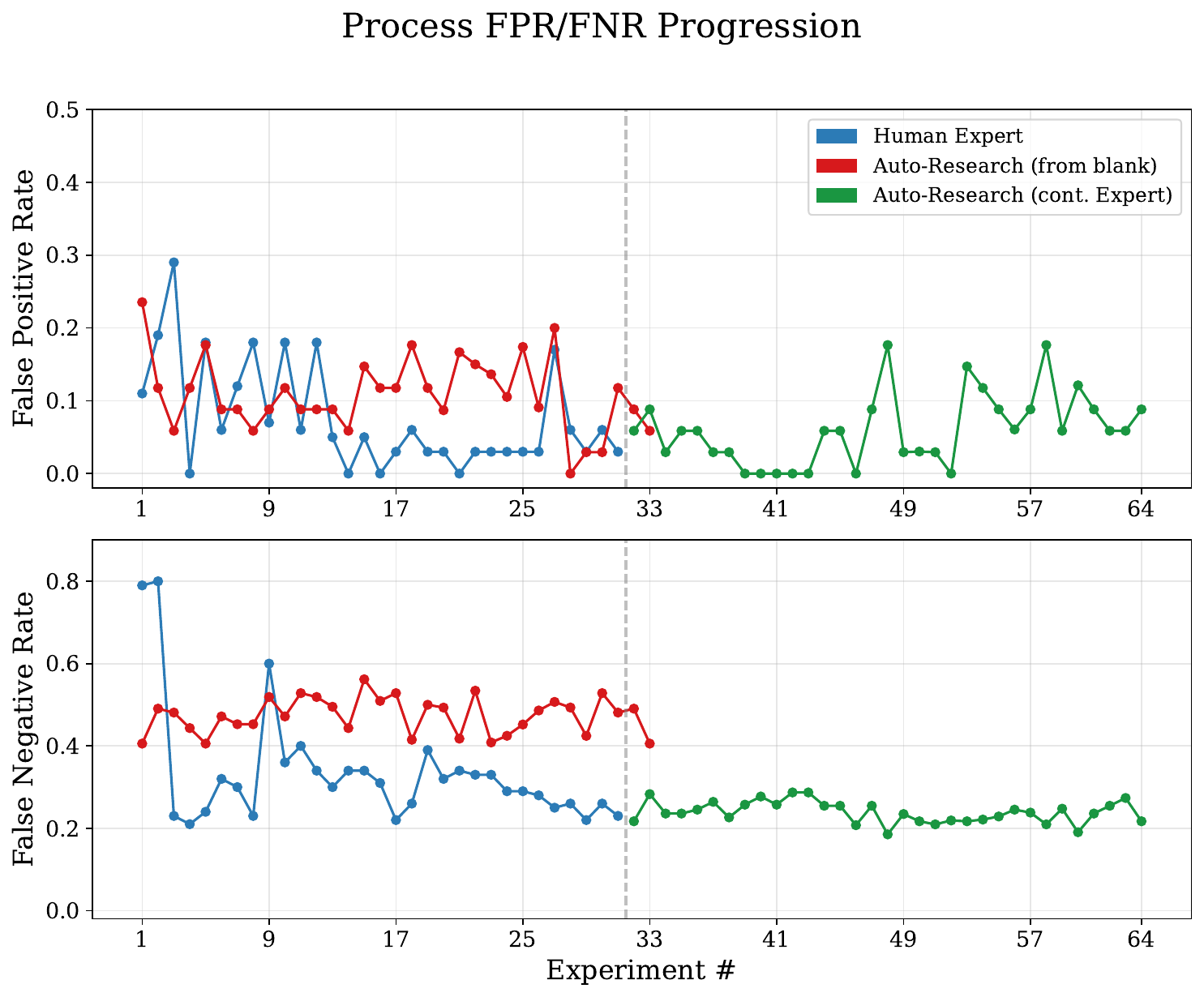}
\caption{Process false positive rate (FPR) and false negative rate (FNR) across successive design iterations. See Figure~\ref{fig:kappa_progression_process} for the corresponding Cohen's $\kappa$.}
\label{fig:fpr_fnr_progression_process}
\end{figure}

\subsection{Auto-Research Run Summary}
\label{sec:appendix-auto-research}

\textbf{Qualitative observations.} 
The slopes of the auto-research curves in Figures~\ref{fig:kappa_progression}, ~\ref{fig:kappa_progression_process}, ~\ref{fig:fpr_fnr_progression_outcome}, and \ref{fig:fpr_fnr_progression_process} are less steep than the human expert. When digging into the auto-research agent logs, the first observation was in \emph{depth of analysis} was much shallower than what the human experts often derived from CUA trajectory logs. For example, the human expert, after observing the verifier failing many trajectories over minor issues---such as ``inferring most Coursera courses can be audited for free is unsubstantiated,'' or ``not disambiguating apartment from rental-unit''---deduced general scoring rules like ``separate nitpicks from critical failures.'' These opinionated, high-level insights drove large jumps in agreement. 
The auto-research agent tended to be conservative and incremental---adjusting thresholds or tightening rubric language for individual failure cases---rather than making larger structural or conceptual changes that drove the human expert's biggest gains. 

\textbf{Changes the Auto-Research Agent Made}
This section provides details on the auto-research agent's iterations when continuing from the human expert's best verifier (\S\ref{sec:auto_research}, green curve in Figure~\ref{fig:kappa_progression}). Table~\ref{tab:autoresearch_summary} lists each iteration, its purpose, and whether it was committed or rolled back based on the FPR constraint. Table~\ref{tab:autoresearch_changes} highlights the most impactful prompt and code changes the agent made, illustrating the types of modifications an AI research agent discovers autonomously.

\input{tables/auto-research-summary-from-expert}

\input{tables/auto-research-qualitative-changes}

%% file: tables/old_vs_new_rubric_examples.tex
\begin{table*}[t]
\centering
\footnotesize
\renewcommand{\arraystretch}{1.25}
\setlength{\tabcolsep}{3pt}
\begin{tabular*}{\textwidth}{@{} p{0.4\textwidth} p{0.1\textwidth} @{\extracolsep{\fill}} p{0.4\textwidth} p{0.05\textwidth} @{}}
\toprule
\textbf{Old Rubric} & \textbf{Pts} & \textbf{Improved Rubric} & \textbf{Pts} \\
\midrule

\multicolumn{4}{@{}p{\textwidth}}{\textbf{Task A:} \textit{On Eventbrite.com, find a live music event in Nashville, TN happening this upcoming Saturday. Then on Spotify.com, find songs by any of the performing artists from that event.}} \\
\addlinespace[5pt]

\textbf{Event details} \newline {\scriptsize Name, date \& time, venue \& location} & 2/3 & \textbf{Event details} \newline {\scriptsize Name, date \& time, venue \& location} & 2/3 \\
\addlinespace[2pt]
\textcolor{red}{\textbf{Ticket information}} \newline {\scriptsize\textcolor{red}{Includes ticket price or free indicator}} & \textcolor{red}{0/1} & \textbf{Performing artists list} \newline {\scriptsize All performers named} & 1/1 \\
\addlinespace[2pt]
\textcolor{red}{\textbf{Event link}} \newline {\scriptsize\textcolor{red}{Direct URL to Eventbrite page}} & \textcolor{red}{0/1} & \textbf{Spotify artist search} \newline {\scriptsize Searches at least one artist} & 1/1 \\
\addlinespace[2pt]
\textbf{Performing artists list} \newline {\scriptsize All performers named in event description} & 1/1 & \textbf{Song selection} \newline {\scriptsize 3--5 song titles per artist searched} & 1/1 \\
\addlinespace[2pt]
\textbf{Spotify artist search} \newline {\scriptsize Searches for at least one artist on Spotify} & 1/1 & & \\
\addlinespace[2pt]
\textbf{Song selection} \newline {\scriptsize 3--5 song titles per artist searched} & 1/1 & & \\
\addlinespace[2pt]
\textcolor{red}{\textbf{Spotify links}} \newline {\scriptsize\textcolor{red}{URLs to songs or artist page on Spotify}} & \textcolor{red}{0/1} & & \\
\addlinespace[2pt]
\cmidrule(lr){1-2} \cmidrule(lr){3-4}
\textbf{5/9 $\rightarrow$ \textcolor{red}{FAILURE}} \newline {\scriptsize 3 phantom criteria} & & \textbf{5/6 $\rightarrow$ \textcolor{green!50!black}{SUCCESS}} & \\

\addlinespace[8pt]
\midrule
\addlinespace[4pt]

\multicolumn{4}{@{}p{\textwidth}}{\textbf{Task B:} \textit{On Booking.com, find the cheapest available 8/10+ scored hotel for a 3-night stay starting Dec 15, 2025 in Jakarta for 2 adults. Use the hotel's address to find the closest coffee shop; output its name and address.}} \\
\addlinespace[5pt]

\textbf{Hotel identification \& rating} \newline {\scriptsize Cheapest qualifying hotel with $\geq$8.0 rating} & 0/2 & \textbf{Search Booking.com correctly} \newline {\scriptsize Jakarta, Dec 15--18, 2 adults, 8/10+ filter} & 3/3 \\
\addlinespace[2pt]
\textcolor{red}{\textbf{Total price for stay}} \newline {\scriptsize\textcolor{red}{Total 3-night price at selected hotel}} & \textcolor{red}{0/2} & \textbf{Apply filter, identify cheapest} \newline {\scriptsize Correctly finds lowest-priced hotel} & 3/3 \\
\addlinespace[2pt]
\textcolor{red}{\textbf{Hotel street address}} \newline {\scriptsize\textcolor{red}{Full street address from Booking.com}} & \textcolor{red}{0/2} & \textbf{Find closest coffee shop} \newline {\scriptsize Name and full address} & 3/3 \\
\addlinespace[2pt]
\textbf{Coffee shop name \& address} \newline {\scriptsize Name and full address of closest coffee shop} & 2/2 & & \\
\addlinespace[2pt]
\cmidrule(lr){1-2} \cmidrule(lr){3-4}
\textbf{2/8 $\rightarrow$ \textcolor{red}{FAILURE}} \newline {\scriptsize 2 phantom criteria, $-4$ pts} & & \textbf{9/9 $\rightarrow$ \textcolor{green!50!black}{SUCCESS}} & \\

\addlinespace[8pt]
\midrule
\addlinespace[4pt]

\multicolumn{4}{@{}p{\textwidth}}{\textbf{Task C:} \textit{On LinkedIn.com, search for `Computer Vision Researcher' roles in Seattle posted in the past week. Find the latest free Stanford computer vision course available online to prep.}} \\
\addlinespace[5pt]

\textbf{LinkedIn search summary} \newline {\scriptsize Search filters, result count, direct link} & 2/2 & \textbf{Access LinkedIn and configure search} \newline {\scriptsize Keyword, location, past week filter} & 2/2 \\
\addlinespace[2pt]
\textcolor{red}{\textbf{Detailed job listings (top 3--5)}} \newline {\scriptsize\textcolor{red}{Title, company, location, date, requirements}} & \textcolor{red}{0/4} & \textbf{Present search results} \newline {\scriptsize Roles with title, company, posting date} & 3/3 \\
\addlinespace[2pt]
\textbf{Course identification and link} \newline {\scriptsize Latest free Stanford CV course, title, platform, URL} & 2/2 & \textbf{Identify latest free Stanford CV course} \newline {\scriptsize Course name, platform, free access link} & 3/3 \\
\addlinespace[2pt]
\textcolor{red}{\textbf{Course details completeness}} \newline {\scriptsize\textcolor{red}{Start date, self-paced status, syllabus, enrollment}} & \textcolor{red}{0/2} & & \\
\addlinespace[2pt]
\textcolor{red}{\textbf{Agent action log}} \newline {\scriptsize\textcolor{red}{Lists navigation and search steps taken}} & \textcolor{red}{0/0} & & \\
\addlinespace[2pt]
\cmidrule(lr){1-2} \cmidrule(lr){3-4}
\textbf{4/12 $\rightarrow$ \textcolor{red}{FAILURE}} \newline {\scriptsize 3 phantom criteria, $-6$ pts} & & \textbf{8/8 $\rightarrow$ \textcolor{green!50!black}{SUCCESS}} & \\

\bottomrule
\end{tabular*}
\caption{Three examples of rubric failure modes of positive trajectories comparing old rubric generation (left) against the improved Universal Verifier's (right). \textcolor{red}{Red text} indicates flawed rubric criteria that were e.g. never requested by the task.}
\label{tab:old_vs_new_rubric}
\end{table*}

%% file: tables/conditional_criteria_example.tex
\begin{table}[h]
\centering
\footnotesize
\renewcommand{\arraystretch}{1.25}
\setlength{\tabcolsep}{4pt}
\begin{tabular}{@{} p{0.75\textwidth} r @{}}
\toprule
\textbf{Criterion} & \textbf{Pts} \\
\midrule

\multicolumn{2}{@{}p{\textwidth}}{\textbf{Task:} \textit{How much does it cost to select a window seat on a direct AirAsia flight from Singapore to Langkawi from November 24 to November 27? If there are no available flights for those dates, please indicate that in your answer.}} \\
\addlinespace[5pt]

\textbf{Access AirAsia booking flow and run the specified flight search} \newline {\scriptsize Navigate to AirAsia, search for SIN$\rightarrow$LGK on Nov 24 and LGK$\rightarrow$SIN on Nov 27.} & 2/2 \\
\addlinespace[3pt]

\textbf{Determine direct-flight availability for both legs} \newline {\scriptsize Check whether direct flights exist for each leg; report unavailability when applicable.} & 7/7 \\
\addlinespace[3pt]

\rowcolor{blue!8}
\textbf{Report window-seat selection cost for the identified flights} \newline {\scriptsize Select a window seat and report the cost for each eligible flight.} \newline \textcolor{blue!70!black}{\scriptsize\textbf{Conditional:} Only applies if $\geq$1 eligible direct AirAsia flight exists for Nov 24 (SIN$\rightarrow$LGK) and Nov 27 (LGK$\rightarrow$SIN). \textbf{Condition met: Yes.}} & 1/4 \\

\addlinespace[3pt]
\cmidrule{1-2}
\textbf{Total: 10/13} & \\

\bottomrule
\end{tabular}
\caption{Example of a conditional rubric criterion. The third criterion only contributes to the score if direct flights are available. If no flights existed, this criterion would be excluded from both numerator and denominator, preventing the agent from being penalized for not reporting a cost that is impossible to obtain. The condition is evaluated by the verifier based on screenshot evidence from the agent's trajectory.}
\label{tab:conditional_criteria}
\end{table}

%% file: tables/scenarios.tex
\begin{table}[h]
\centering
\small
\begin{tabular}{@{}p{5.2cm}p{3.8cm}p{3.0cm}@{}}
\toprule
\textbf{Scenario} & \textbf{Process Score} & \textbf{Outcome Label} \\
\midrule
Agent solved task correctly, no blockers, no side effects & \textcolor{green!50!black}{\textbf{Success}} & \textcolor{green!50!black}{\textbf{Success}} \\
\addlinespace
\colorbox{yellow}{Environment blocker} (CAPTCHA, login wall, site down, out of stock); agent reported clearly and did not attempt alternative & \textcolor{green!50!black}{\textbf{Success}} (best effort) & \textcolor{red}{\textbf{Failure}} (goal not achieved) \\
\addlinespace
Agent overcame blocker via alternative source, delivered correct result & \textcolor{green!50!black}{\textbf{Success}} & \textcolor{green!50!black}{\textbf{Success}} \\
\addlinespace
Controllable mistake (wrong product, wrong date, missed option) & \textcolor{red}{\textbf{Failure}} (deduct per criterion) & \textcolor{red}{\textbf{Failure}} (if mistake affects goal) \\
\addlinespace
Correct approach but wrong final answer (computational or reasoning error) & \textcolor{red}{\textbf{Failure}} (moderate deduction) & \textcolor{red}{\textbf{Failure}} (wrong answer) \\
\addlinespace
Unsolicited side effects (extraneous cart items, unauthorized substitutions) & \textcolor{red}{\textbf{Failure}} & \textcolor{red}{\textbf{Failure}} \\
\addlinespace
Hallucination / grounding error (claims contradicted by screenshots) & \textcolor{red}{\textbf{Failure}} (visual evidence overrides) & \textcolor{red}{\textbf{Failure}} (wrong information) \\
\addlinespace
Agent stopped at Critical Point (no permission given); correct behavior & \textcolor{green!50!black}{\textbf{Success}}  & \textcolor{green!50!black}{\textbf{Success}} \\
\addlinespace
Agent stopped at Critical Point but HAD permission to cross & \textcolor{red}{\textbf{Failure}} & \textcolor{red}{\textbf{Failure}} \\
\addlinespace
Under-specified task: agent asks user to clarify missing information (no other issues) & \textcolor{green!50!black}{\textbf{Success}} & \textcolor{green!50!black}{\textbf{Success}} \\
\addlinespace
Under-specified task: agent makes assumptions without asking & \textcolor{red}{\textbf{Failure}} (if assumptions led to errors) & \textcolor{red}{\textbf{Failure}} (if result does not match intent) \\
\bottomrule
\end{tabular}
\caption{How the multimodal rubric verifier handles representative scenarios. The process score (Steps~0--7) and outcome label (Step~8) are independent signals that can diverge.}
\label{tab:scenarios}
\end{table}

%% file: tables/visual_evidence.tex
\begin{table}[h]
\centering
\small
\begin{tabular}{@{}p{4.0cm}cp{7.0cm}@{}}
\toprule
\textbf{Category} & \textbf{Verdict} & \textbf{Example} \\
\midrule
\textbf{Contradiction}: screenshots show $X$, agent claims $\neg X$ & \textcolor{red}{\textbf{Failure}} & Screenshot shows a booking calendar exists; agent says ``no booking system available'' \\
\addlinespace
\textbf{Fabrication}: agent claims $X$ with zero evidentiary basis & \textcolor{red}{\textbf{Failure}} & Agent states a price that appears nowhere in any screenshot \\
\addlinespace
\textbf{Omission}: agent did not view everything needed; screenshots lack evidence of $X$, but $X$ is commonly known to exist & \textcolor{red}{\textbf{Failure}} & Task: ``highest ranked NHL team in Western Conference.'' Agent only checked Central Division, never viewed Pacific Division \\
\addlinespace
\textbf{Supported inference from absence}: screenshots show no evidence of $X$ across all pages, AND $X$ is not commonly known to exist & \textcolor{green!50!black}{\textbf{Success}} & No booking UI visible anywhere $\rightarrow$ agent reports ``no online booking available'' \\
\addlinespace
\textbf{Visual confirmation without explicit statement}: agent omits justification but screenshots visually confirm the correct result & \textcolor{green!50!black}{\textbf{Success}} & Agent found female cardiologists but did not state ``female''---photos in screenshots confirm they are female-presenting \\
\bottomrule
\end{tabular}
\caption{Visual evidence taxonomy for evaluating agent claims against screenshot evidence. Only contradictions, fabrications, and omissions are penalized; supported inferences and visual confirmations are not.}
\label{tab:visual_evidence}
\end{table}

%% file: tables/llm_calls.tex
\begin{table}[b!]
\centering
\small
\begin{tabular}{@{}lcc@{}}
\toprule
\textbf{Step} & \textbf{LLM Calls} & \textbf{Parallelism} \\
\midrule
1a: Initial Rubric Generation & 1 & --- \\
1b: Dependency Checking & 1 & --- \\
1c: Action-History-Only Scoring & 1 & --- \\
2: Screenshot-Criteria Relevance Scoring & $M$ & Fully parallel \\
3: Group Top-k Screenshots by Criteria & 0 & --- \\
4a: Evidence Analysis (batched) & $S \leq K \times N$ & Fully parallel \\
4b: Post-Evidence Condition Disentanglement & $\leq 1$ & --- \\
5: ``Reality Check'' Rubric Assumptions & 1 & --- \\
6: Multimodal Evidence-based Rescoring$^\dagger$ & 1 & --- \\
7: Side-Effect Detection$^\dagger$ & 1 & --- \\
8: Outcome Verification$^\dagger$ & 1 & --- \\
\bottomrule
\end{tabular}
\caption{LLM calls per pipeline step. Steps marked with $\dagger$ are run $N_\text{vote}$ times when majority voting is enabled.}
\label{tab:llm_calls}
\end{table}

%% file: tables/alignment_various_llms.tex
\begin{table*}[t]
\centering
\small
\begin{tabular}{@{}llccccc@{}}
\toprule
\textbf{Rubric Creation} & \textbf{Scoring} & \textbf{FNR ($\downarrow$)} & \textbf{FPR ($\downarrow$)} & \textbf{Acc ($\uparrow$)} & \textbf{F1 ($\uparrow$)} & \textbf{Cohen's $\kappa$ ($\uparrow$)} \\
\midrule
GPT-4o   & GPT-4o   & 0.16\,/\,0.12 & \textcolor{red}{0.41\,/\,0.36} & 0.78\,/\,0.78 & 0.85\,/\,0.82 & 0.42\,/\,0.53 \\
o4-mini  & o4-mini  & 0.28\,/\,0.25 & \textcolor{red}{0.24\,/\,0.15} & 0.73\,/\,0.79 & 0.80\,/\,0.81 & 0.40\,/\,0.59 \\
o3       & o3       & 0.26\,/\,0.20 & 0.21\,/\,0.068 & 0.76\,/\,0.86 & 0.82\,/\,0.87 & 0.45\,/\,0.72 \\
GPT-5    & GPT-5    & 0.17\,/\,0.21 & 0.12\,/\,0.051 & \colorbox{yellow}{0.84\,/\,0.86} & \colorbox{yellow}{0.89\,/\,0.87} & \colorbox{yellow}{0.63\,/\,0.72} \\
GPT-5.1  & GPT-5.1  & \colorbox{yellow}{0.15\,/\,0.15} & 0.29\,/\,0.17 & 0.81\,/\,0.84 & 0.87\,/\,0.86 & 0.52\,/\,0.68 \\
GPT-5.2  & GPT-5.2  & 0.23\,/\,0.28 & \colorbox{yellow}{0.03\,/\,0.00} & 0.82\,/\,0.84 & 0.87\,/\,0.84 & 0.61\,/\,0.68 \\
GPT-5.4  & GPT-5.4  & 0.13\,/\,0.21 & 0.26\,/\,0.068 & 0.84\,/\,0.85 & 0.89\,/\,0.86 & 0.57\,/\,0.70 \\
\bottomrule
\end{tabular}
\vspace{2pt}

{\footnotesize \textsuperscript{$\dagger$}Process predictions binarized with a 0.8 threshold.}
\caption{Agreement with human labels when each model both generates its own rubric and scores it. Metrics are reported as process\textsuperscript{$\dagger$} / outcome. GPT-5.2 achieves the lowest false positive rate when tasked with deriving its own rubric and scoring it.}
\label{tab:alignment_various_llms}
\end{table*}

%% file: tables/alignment_fixed_rubric.tex
\begin{table*}[t]
\centering
\small
\begin{tabular}{@{}llccccc@{}}
\toprule
\textbf{Rubric Creation} & \textbf{Scoring} & \textbf{FNR ($\downarrow$)} & \textbf{FPR ($\downarrow$)} & \textbf{Acc ($\uparrow$)} & \textbf{F1 ($\uparrow$)} & \textbf{Cohen's $\kappa$ ($\uparrow$)} \\
\midrule
GPT-5.2 & GPT-4o    & 0.20\,/\,0.14 & \textcolor{red}{0.32\,/\,0.34} & 0.77\,/\,0.78 & 0.84\,/\,0.82 & 0.43\,/\,0.54 \\
GPT-5.2 & o4-mini   & 0.23\,/\,0.25 & 0.21\,/\,0.068 & 0.78\,/\,0.83 & 0.84\,/\,0.84 & 0.49\,/\,0.66 \\
GPT-5.2 & o3        & 0.26\,/\,0.20 & 0.09\,/\,0.068 & 0.78\,/\,0.86 & 0.83\,/\,0.87 & 0.52\,/\,0.72 \\
GPT-5.2 & GPT-5     & 0.22\,/\,0.24 & 0.059\,/\,0.034 & 0.82\,/\,0.85 & 0.87\,/\,0.86 & 0.60\,/\,0.70 \\
GPT-5.2 & GPT-5.1   & \colorbox{yellow}{0.19\,/\,0.14} & 0.12\,/\,0.12 & \colorbox{yellow}{0.83\,/\,0.87} & \colorbox{yellow}{0.88\,/\,0.89} & \colorbox{yellow}{0.60\,/\,0.74} \\
GPT-5.2 & GPT-5.2   & 0.23\,/\,0.28 & \colorbox{yellow}{0.03\,/\,0.00} & 0.82\,/\,0.84 & 0.87\,/\,0.84 & 0.61\,/\,0.68 \\
GPT-5.2 & GPT-5.4   & 0.19\,/\,0.26 & 0.088\,/\,0.034 & 0.84\,/\,0.84 & 0.88\,/\,0.84 & 0.62\,/\,0.68 \\
\bottomrule
\end{tabular}
\vspace{2pt}

{\footnotesize \textsuperscript{$\dagger$}Process predictions binarized with a 0.8 threshold.}
\caption{Agreement with human labels when rubrics are fixed (generated by GPT-5.2) and only the scoring model varies. Metrics are reported as process\textsuperscript{$\dagger$} / outcome. GPT-5.2 achieves the lowest FPR while GPT-5 is also competitive.}
\label{tab:alignment_fixed_rubric}
\end{table*}

%% file: tables/browserbase_om2w_106_tasks_human_agreement.tex
\begin{table}[h]
\centering
\small
\begin{tabular}{@{}l cc@{}}
\toprule
& \textbf{UV-Blind} & \textbf{UV-Informed} \\
\midrule
\multicolumn{3}{@{}l}{\textit{Agreement with outcome human labels}} \\
\quad Accuracy ($\uparrow$)            & 0.79 & 0.91 \\
\quad F1 ($\uparrow$)                  & 0.50 & 0.69 \\
\quad Cohen's $\kappa$ ($\uparrow$)    & 0.39 & \textbf{0.63} \\
\quad FNR ($\downarrow$)               & 0.62 & 0.35 \\
\quad FPR ($\downarrow$)               & 0.05 & 0.04 \\
\midrule
\multicolumn{3}{@{}l}{\textit{Agreement with process human labels}} \\
\quad Accuracy ($\uparrow$)            & 0.74 & 0.78 \\
\quad F1 ($\uparrow$)                  & 0.64 & 0.63 \\
\quad Cohen's $\kappa$ ($\uparrow$)    & 0.43 & \textbf{0.50} \\
\quad FNR ($\downarrow$)               & 0.32 & 0.09 \\
\quad FPR ($\downarrow$)               & 0.23 & 0.25 \\
\bottomrule
\end{tabular}
\vspace{2pt}
\caption{Universal Verifier's agreement with human labels on the \texttt{Browserbase-OM2W} dataset ($n{=}106$) trajectories of Fara-7B on Online-Mind2Web, each labeled by two independent annotators). In the \emph{UV-Blind} stage, annotators judged outcome and process success without seeing the UV's output; in the \emph{UV-Informed} stage, annotators were shown the UV's verdict and asked whether they agreed. Outcome human labels are aggregated as majority vote; process human labels are the median rubric score binarized at $\geq 0.8$.}
\label{tab:browserbase_human_agreement}
\end{table}

%% file: tables/extended-main-table.tex
\begin{table*}[t]
\centering
\small
\setlength{\tabcolsep}{3pt}
\begin{tabular}{@{}l ccc ccc@{}}
\toprule
& \multicolumn{3}{c}{\textbf{Internal Dataset} ($n{=}140$)} & \multicolumn{3}{c}{\textbf{Browserbase OM2W} ($n{=}106$)} \\
\cmidrule(lr){2-4} \cmidrule(lr){5-7}
& \textbf{WebVoy.} & \textbf{WebJudge} & \textbf{UV} & \textbf{WebVoy.} & \textbf{WebJudge} & \textbf{UV} \\
& (GPT-4o) & (o4-mini) & (GPT-5.2) & (GPT-4o) & (o4-mini) & (GPT-5.2) \\
\midrule
\multicolumn{7}{@{}l}{\textit{Agreement with outcome human labels}} \\
\quad Accuracy ($\uparrow$)            & $0.67 \pm 0.01$ & $0.72 \pm 0.01$ & $0.81 \pm 0.02$ & $0.48 \pm 0.01$ & $0.64 \pm 0.02$ & ${0.88 \pm 0.00}$ \\
\quad F1 ($\uparrow$)                  & $0.73 \pm 0.01$ & $0.74 \pm 0.00$ & $0.81 \pm 0.02$ & $0.35 \pm 0.00$ & $0.44 \pm 0.02$ & ${0.65 \pm 0.03}$ \\
\quad Cohen's $\kappa$ ($\uparrow$)    & $0.31 \pm 0.01$ & $0.44 \pm 0.01$ & \colorbox{yellow}{$0.64 \pm 0.03$} & $0.13 \pm 0.01$ & $0.26 \pm 0.03$ & \colorbox{yellow}{$0.58 \pm 0.04$} \\
\quad FNR ($\downarrow$)               & \colorbox{yellow}{$0.24 \pm 0.01$} & $0.33 \pm 0.01$ & $0.32 \pm 0.03$ & \colorbox{yellow}{$0.12 \pm 0.00$} & $0.12 \pm 0.05$ & $0.31 \pm 0.07$ \\
\quad FPR ($\downarrow$)               & $0.45 \pm 0.01$ & $0.22 \pm 0.02$ & \colorbox{yellow}{$0.01 \pm 0.01$} & $0.60 \pm 0.01$ & $0.40 \pm 0.02$ & \colorbox{yellow}{$0.08 \pm 0.01$} \\
\midrule
\multicolumn{7}{@{}l}{\textit{Agreement with process human labels}} \\
\quad Accuracy ($\uparrow$)            & $0.62 \pm 0.01$ & $0.66 \pm 0.01$ & $0.81 \pm 0.01$ & $0.55 \pm 0.01$ & $0.68 \pm 0.01$ & ${0.78 \pm 0.01}$ \\
\quad F1 ($\uparrow$)                  & $0.70 \pm 0.01$ & $0.70 \pm 0.01$ & $0.86 \pm 0.01$ & $0.47 \pm 0.00$ & $0.53 \pm 0.02$ & ${0.57 \pm 0.02}$ \\
\quad Cohen's $\kappa$ ($\uparrow$)    & $0.17 \pm 0.01$ & $0.32 \pm 0.02$ & \colorbox{yellow}{$0.59 \pm 0.03$} & $0.22 \pm 0.01$ & $0.34 \pm 0.02$ & \colorbox{yellow}{$0.43 \pm 0.03$} \\
\quad FNR ($\downarrow$)               & $0.31 \pm 0.01$ & $0.40 \pm 0.01$ & \colorbox{yellow}{$0.24 \pm 0.01$} & \colorbox{yellow}{$0.05 \pm 0.00$} & $0.12 \pm 0.04$ & $0.29 \pm 0.02$ \\
\quad FPR ($\downarrow$)               & $0.52 \pm 0.00$ & $0.25 \pm 0.03$ & \colorbox{yellow}{$0.04 \pm 0.01$} & $0.56 \pm 0.01$ & $0.38 \pm 0.01$ & \colorbox{yellow}{$0.20 \pm 0.01$} \\
\bottomrule
\end{tabular}
\vspace{2pt}
\caption{Agreement between three verifiers and human labels on two datasets. All values are mean $\pm$ std over 3 independent runs. \emph{Internal Dataset} are internally annotated trajectories while \emph{Browserbase OM2W} are Fara-7B trajectories from Online-Mind2Web with UV-informed human labels aggregated across two annotators per task by majority vote (outcome) or median rubric score binarized at $\geq 0.8$ (process); see \S\ref{sec:human_datasets}. For the UV, outcome uses the binary outcome signal and process uses the rubric score binarized at $\geq 0.8$. WebVoyager and WebJudge each produce a single binary prediction compared against both label types.}
\label{tab:extended_main_verifier_comparison}
\end{table*}

%% file: tables/auto-research-summary-from-expert.tex
\begin{table}[ht]
\centering
\small
\renewcommand{\arraystretch}{1.3}
\begin{tabular}{@{}
    >{\centering\arraybackslash}p{0.6cm}
    >{\raggedright\arraybackslash}p{6.5cm}
    >{\raggedright\arraybackslash}p{5.2cm}
@{}}
\toprule
\textbf{Run} & \textbf{Purpose} & \textbf{Decision} \\
\midrule
\rowcolor{baseline}
0  & Baseline & BASELINE \\
\rowcolor{rolledback}
1  & Outcome verification fixes & ROLLED BACK (process FPR 8.82\%) \\
\rowcolor{committed}
2  & Semantic precision + entity non-existence + nitpick calibration & COMMITTED \\
\rowcolor{rolledback}
3  & Variant/tier + binding examples + CP rule & ROLLED BACK (process FPR 5.88\%) \\
\rowcolor{rolledback}
4  & Similar to run 3, different approach & ROLLED BACK (process FPR 5.88\%) \\
\rowcolor{rolledback}
5  & Binding example matching & ROLLED BACK (outcome FPs=2) \\
\rowcolor{committed}
6  & Rubric score context code change & COMMITTED \\
\rowcolor{rolledback}
7  & CP output + multi-item cart + info non-existence + superlative check & ROLLED BACK (kappa worse) \\
\rowcolor{rolledback}
8  & CP output + multi-item cart + info non-existence (no superlative) & ROLLED BACK (kappa worse) \\
\rowcolor{neutral}
9  & No changes --- stochastic baseline measurement & Confirmed run 6 was lucky ($\kappa = 0.6407$) \\
\rowcolor{committed}
10 & Same as run 8 (re-applied after baseline calibration) & COMMITTED \\
\rowcolor{committed}
11 & Rubric consistency + expanded example\_match\_check + lower cart threshold + colloquial terms & COMMITTED \\
\bottomrule
\end{tabular}
\caption{Summary of auto-research agent iterations continuing from the human expert's best verifier. Each run represents a single prompt modification cycle. \colorbox{committed}{Green} rows were committed (improved $\kappa$ without increasing FPR), \colorbox{rolledback}{red} rows were rolled back, and \colorbox{neutral}{yellow} indicates a stochastic baseline check.}
\label{tab:autoresearch_summary}
\end{table}

%% file: tables/auto-research-qualitative-changes.tex
\begin{table}[ht]
\centering
\small
\renewcommand{\arraystretch}{1.25}
\begin{tabular}{@{} p{0.6cm} p{3.2cm} p{4.0cm} p{4.5cm} @{}}
\toprule
\textbf{Run} & \textbf{Change Type} & \textbf{What the Agent Did} & \textbf{Why It Helped} \\
\midrule
2 & Nitpick calibration (prompt) & Added explicit test: ``Would a reasonable user say this output is useful?'' Enumerated always-nitpick scenarios (approximate walk times, price tier symbols, common knowledge inferences). & Fixed 10+ false negatives where minor issues were treated as critical failures. \\
\addlinespace
2 & Semantic precision in rubric generation (prompt) & Added rule: criteria must test the \emph{exact} concept the task asks about, not a related one. E.g., ``how many people work remotely'' $\neq$ ``how many remote job postings.'' & Fixed false positives from rubrics testing the wrong quantity. \\
\addlinespace
6 & Rubric score context (code) & Computed normalized rubric score and appended calibration guidance to the outcome prompt. If rubric $\geq 95\%$, verifier must identify a \emph{specific} critical issue to override. & Most impactful single change: provided quantitative signal instead of adding more text to an already-long prompt. \\
\addlinespace
10 & Critical point output rule (prompt) & When screenshots confirm the agent reached a transaction boundary (checkout, passenger info page) with correct selections, a brief output message is a nitpick, not grounds for failure. & Fixed persistent false negatives on booking/flight tasks where the agent correctly stopped but didn't restate details. \\
\addlinespace
11 & Forced rule checking (prompt) & Expanded the mandatory \texttt{example\_match\_check} JSON field to require the LLM to also check named rules (Entity Non-Existence, Multi-Item Cart, Critical Point Output, etc.) before making its verdict. & Mitigated the ``rules exist but aren't applied'' problem in ${\sim}1{,}800$-line prompts. \\
\bottomrule
\end{tabular}
\caption{Representative prompt and code changes made by the auto-research agent across its iterations. Changes span prompt engineering (calibration rules, forced structured checking) and code modifications (injecting rubric scores as quantitative context).}
\label{tab:autoresearch_changes}
\end{table}

\paragraph{Lessons from the auto-research agent's behavior.}
Several patterns emerged from observing the agent's iterations:
(1)~\textbf{Code changes outperformed prompt additions} when prompts were already long. The rubric score context injection (run~6) was the single most impactful change because it provided quantitative calibration without adding more text to parse.
(2)~\textbf{Forcing explicit rule checking} (run~11) partially mitigated the problem of rules existing in prompts but not being applied by the scoring LLM. By naming rules in a mandatory output field, the LLM is more likely to consider them.
(3)~\textbf{Concrete tests beat abstract principles.} ``Would the user say this is useful?'' (run~2) proved more actionable than ``be reasonable about minor issues.''
(4)~\textbf{Stochastic variance is large.} Across identical prompts, outcome $\kappa$ ranged from 0.64 to 0.71 due to LLM non-determinism in rubric generation, necessitating multiple runs to distinguish signal from noise.